\documentclass[draftcls, onecolumn, 12 pt]{IEEEtran}

\newcommand{\unicode}[1]{{}}

\usepackage{cite}
\usepackage{graphicx}
\usepackage{epsfig}
\usepackage{amsmath}
\usepackage{amsfonts}
\usepackage{psfrag}
\usepackage{rotating}
\usepackage{latexsym} 
\usepackage{amsthm}
\usepackage{amssymb}
\usepackage{amsmath}
\usepackage{fancyvrb}
\usepackage{textcomp}
\usepackage{url}
\usepackage{ifdraft}
\usepackage{multirow}
\usepackage{rotating}
\usepackage{xspace}
\usepackage{array}
\usepackage[usenames]{color}
\usepackage{arydshln}
\usepackage{multido}
\usepackage{enumerate}
\usepackage[latin1]{inputenc}
\usepackage{color}
\usepackage{ifsym}
\usepackage{tipa}
\usepackage{graphics}
\usepackage{epstopdf}
\usepackage{amsfonts}
\usepackage{flushend}
\usepackage{lscape}
\usepackage{pdflscape}
\usepackage[caption = false]{subfig}

\theoremstyle{plain}

\theoremstyle{plain}

\theoremstyle{plain}

\theoremstyle{plain}

\newtheoremstyle{specialcasestyle}
  {5mm}					
  {5mm}					
  {\upshape}			
  {}					
  {\bfseries\upshape}	
  {.}					
  {1mm}					
  {}					
\theoremstyle{specialcasestyle}

\newcommand{\figref}[1]{Fig.~\protect\ref{#1}}

\newcommand{\tableref}[1]{{Table~\protect\ref{#1}}}

\def\suscript(#1,#2,#3){{#1}^{#2}_{#3}}
\newcommand{\fracparams}[2]{\genfrac{}{}{0pt}{}{{#1}}{{#2}}}

\newcommand{\FoxH}[6][right]{
	\ifthenelse{\equal{#1}{right}}{\suscript(\rm{H},{#2},{#3}){\left[{#4}\left|\fracparams{#5}{#6}\right.\right]}}{
		\ifthenelse{\equal{#1}{left}}{\suscript(\rm{H},{#2},{#3}){\left[\left.{#4}\right|\fracparams{#5}{#6}\right]}}{
			\suscript(\rm{H},{#2},{#3}){\left[{#4}\left|\fracparams{#5}{#6}\right.\right]}
		}
	}
}

\newcommand{\FoxHBar}[6][right]{
	\ifthenelse{\equal{#1}{right}}{\suscript(\rm{\bar{H}},{#2},{#3}){\left[{#4}\left|\fracparams{#5}{#6}\right.\right]}}{
		\ifthenelse{\equal{#1}{left}}{\suscript(\rm{\bar{H}},{#2},{#3}){\left[\left.{#4}\right|\fracparams{#5}{#6}\right]}}{
			\suscript(\rm{\bar{H}},{#2},{#3}){\left[{#4}\left|\fracparams{#5}{#6}\right.\right]}
		}
	}
}

\newcommand{\ExtendedFoxHBar}[6][right]{
	\ifthenelse{\equal{#1}{right}}{\suscript(\rm{\hat{H}},{#2},{#3}){\left[{#4}\left|\fracparams{#5}{#6}\right.\right]}}{
		\ifthenelse{\equal{#1}{left}}{\suscript(\rm{\hat{H}},{#2},{#3}){\left[\left.{#4}\right|\fracparams{#5}{#6}\right]}}{
			\suscript(\rm{\hat{H}},{#2},{#3}){\left[{#4}\left|\fracparams{#5}{#6}\right.\right]}
		}
	}
}

\newcommand{\MeijerG}[6][right]{
	\ifthenelse{\equal{#1}{right}}{\suscript(\rm{G},{#2},{#3}){\left[{#4}\left|\fracparams{#5}{#6}\right.\right]}}{
		\ifthenelse{\equal{#1}{left}}{\suscript(\rm{G},{#2},{#3}){\left[\left.{#4}\right|\fracparams{#5}{#6}\right]}}{
			\suscript(\rm{G},{#2},{#3}){\left[{#4}\left|\fracparams{#5}{#6}\right.\right]}
		}
	}
}
\newcommand{\Hypergeom}[5]{
	\suscript({},{},{#1})\suscript({F},{},{#2})\left[{#3};{#4};{#5} \right]
}

\newcommand{\BesselI}[2][0]{
	\suscript({I},{},{#1})\left({#2}\right)
}
\newcommand{\BesselK}[2][0]{
	\suscript({K},{},{#1})\left({#2}\right)
}

\newcommand{\Expected}[1]{
	{{\mathbb{E}}\left[{#1}\right]}
}

\begin{document}
\title{Ergodic Capacity Analysis of Free-Space Optical Links with Nonzero Boresight Pointing Errors}


\author{Imran~Shafique~Ansari,~\IEEEmembership{Student Member,~IEEE}, Mohamed-Slim~Alouini,~\IEEEmembership{Fellow,~IEEE}, and Julian~Cheng,~\IEEEmembership{Senior Member,~IEEE}

\normalsize
\thanks{Imran~Shafique~Ansari and Mohamed-Slim~Alouini are with the Computer, Electrical, and Mathematical Sciences and Engineering (CEMSE) Division at King Abdullah University of Science and Technology (KAUST), Thuwal, Makkah Province, Kingdom of Saudi Arabia (e-mail:~\{imran.ansari, slim.alouini\}@kaust.edu.sa).}
\thanks{Julian~Cheng is with the School of Engineering, The University of Britich Columbia (UBC), Kelowna, BC, V1V 1V7, Canada (e-mail:~julian.cheng@ubc.ca).}
\thanks{This work was supported by a grant from King Abdul-Aziz City of Science and Technology (KACST) number: AT-34-145.}
\thanks{This work is published in \emph{IEEE Transactions on Wireless Communications}.}
}

\normalsize

\maketitle

\begin{abstract}
A unified capacity analysis of a free-space optical (FSO) link that accounts for nonzero boresight pointing errors and both types of detection techniques (i.e. intensity modulation/direct detection as well as heterodyne detection) is addressed in this work. More specifically, an exact closed-form expression for the moments of the end-to-end signal-to-noise ratio (SNR) of a single link FSO transmission system is presented in terms of well-known elementary functions. Capitalizing on these new moments expressions, we present approximate and simple closed-form results for the ergodic capacity at high and low SNR regimes. All the presented results are verified via computer-based Monte-Carlo simulations.
\end{abstract}


\IEEEpeerreviewmaketitle

\section{Introduction}
\subsection{Background}
\IEEEPARstart{I}{n} recent times, radio frequency (RF) spectrum scarcity has become one of the biggest and prime concern in the arena of wireless communications. Due to this RF spectrum scarcity, additional RF bandwidth allocation, as utilized in the recent past, is not anymore a viable solution to fulfill the demand for higher data rates \cite{haykin}. Of the many other popular solutions, free-space optical (FSO) systems have gained an increasing interest due to their advantages including higher bandwidth and higher capacity compared to the traditional RF communication systems.

FSO links are license-free and hence are cost-effective relative to the traditional RF links. FSO is indeed a promising technology as it offers full-duplex Gigabit Ethernet throughput in certain applications and environment offering a huge license-free spectrum, immunity to interference, and high security \cite{popoola}. These features of FSO communication systems potentially enable solving the issues that the RF communication systems face due to the expensive and scarce spectrum (see \cite{peppas1,peppas4,ansari6} and references therein). Additionally, FSO communications does offer bandwidth as the world record stands at 1.2 Tbps or 1200 Gbps \cite{hogan}. With the correct setup, much higher speeds may be possible as the approach utilizes multiple wavelengths acting like separate channels. Hence, in this concept, the signals are sent down a fiber and launched through the air (known as fiber over the air) and then they travel through a lens before ending up back in fiber \cite{hogan}. Besides these nice characteristic features of FSO communication systems, they span over long distances of 1Km or longer. However, the atmospheric turbulence may lead to a significant degradation in the performance of the FSO communication systems \cite{andrews}.

Thermal expansion, dynamic wind loads, and weak earthquakes result in the building sway phenomenon that causes vibration of the transmitter beam leading to a misalignment between transmitter and receiver known as pointing error. These pointing errors may lead to significant performance degradation and are a serious issue in urban areas, where the FSO equipments are placed on high-rise buildings \cite{sandalidis2}. It is worthy to learn that intensity modulation/direct detection (IM/DD) is the main mode of detection in FSO systems but coherent communications have also been proposed as an alternative detection mode. Among these, heterodyne detection is a more complicated detection method but has the ability to better overcome the turbulence effects (see \cite{tsiftsis} and references cited therein).

\subsection{Motivation}
Over the last couple of decades, a good amount of work has been done on studying the performance of a single FSO link operating over weak turbulence channels modeled by lognormal (LN) distribution (see \cite{fried,niu2,zhu1,cheng,liu3,ansari13} and references cited therein), operating over composite turbulence channels (such as Rician-lognormal (RLN) (see \cite{yang2,churnside,song1,yang3,ansari13} and references cited therein)), and operating over generalized turbulence channels modeled by M\'{a}laga ($\mathcal{M}$) distribution (see \cite{navas,balsells,wang2,ansari12} and references therein) and Gamma-Gamma (GG) distribution (as a special case to $\mathcal{M}$ distribution) (see \cite{andrews,peppas1,popoola,park,safari,navidpour,kedar,ansari6,liu,sandalidis3,ansari11} and references therein) under heterodyne detection as well as IM/DD techniques. However, as per authors' best knowledge, there are no unified exact expressions nor asymptotic expressions that capture the ergodic capacity performance of both these detection techniques with nonzero boresight pointing errors under such turbulence channels.

\subsection{Contributions}
The key contributions of this work are stated as follows.
\begin{itemize}
\item The integrals are setup for the ergodic capacity of the LN, the RLN, and the $\mathcal{M}$ (also GG as a special case of $\mathcal{M}$) turbulence models in composition with nonzero boresight pointing errors. On analyzing these integrals, it is realized that most of these \textit{integrals are very complex to solve} and to the authors' best knowledge, an exact closed-form solution to most of these integrals is not achievable. Hence, it is required to look into alternative solutions to analyze the ergodic capacity for such turbulence models.
\item A \textit{unified approach for the calculation of the moments} of a single FSO link is presented in exact closed-form in terms of simple elementary functions for the LN, the RLN, and the $\mathcal{M}$ (also GG as a special case of $\mathcal{M}$) turbulence models. These unified moments are then utilized, as an alternative solution, to perform the ergodic capacity analysis for such turbulence models.
\item A \textit{general methodology} is presented for simplifying the ergodic capacity analysis of composite FSO turbulence models by independently integrating the various constituents of the composite turbulence model thereby trying to reduce the number of integrals. If succeeded to reduce to a single integral (that is not solvable further) then various techniques such as Gauss-Hermite formula can be utilized to obtain the required results.
\item \textit{Asymptotic closed-form expressions} for the ergodic capacity of the LN, the RLN, and the $\mathcal{M}$ (also GG as a special case of $\mathcal{M}$) FSO turbulence models, applicable to high as well as low signal-to-noise ratio (SNR) regimes, are derived in terms of simple elementary functions via utilizing the derived unified moments.
\end{itemize}

\subsection{Structure}
The remainder of the paper is organized as follows. Sections II presents the channel and system model inclusive of the nonzero boresight pointing error model and the various turbulence models applicable to both the types of detection techniques (i.e. heterodyne detection and IM/DD) utilized in this work. Section III presents the derivation of the exact closed-form channel statistic in terms of the moments in simple elementary functions for the various turbulence models introduced in Section II under the effects of nonzero boresight effects. Ergodic capacity analysis in terms of approximate though closed-form expressions is presented along with some simulation results to validate these analytical results in Section IV for these turbulence channels in terms of simple elementary functions. Finally Section V makes some concluding remarks.

\section{Channel and System Model}
We consider a FSO system with either of the two types of detection techniques i.e. heterodyne detection (denoted in our formulas by $r=1$) or IM/DD (denoted in our formulas by $r=2$). The transmitted data propagates through an atmospheric turbulence channel in the presence of pointing errors. The received optical power is converted into an electrical signal through either of the two types of detection technique (i.e. heterodyne detection or IM/DD) at the photodetector. Assuming additive white Gaussian noise (AWGN) $N$ for the thermal/shot noise, the received signal $y$ can be expressed as
\begin{equation}
y = I\,x+N,
\end{equation}
where $x$ is the transmit intensity and $I$ is the channel gain. Following \cite{farid,yang3}, we assume that the off-axis scintillation varies slowly near the spot of boresight displacement and uses a constant value of scintillation index to characterize the atmospheric turbulence. Hence, the atmospheric turbulence and the pointing error are independent. Subsequently, the channel gain can be expressed as $I=I_{l}\,I_{a}\,I_{p}$, where $I_{l}$ is the path loss that is a constant in a given weather condition and link distance, $I_{a}$ is a random variable that signifies the atmospheric turbulence loss factor, and $I_{p}$ is another random variable that represents the pointing error loss factor.

\subsection{Pointing Error Models}
\subsubsection{Nonzero Boresight Pointing Error Model}
Pointing error impairments are assumed and employed to be present for which the probability density function (PDF) of the irradiance $I_{p}$ with nonzero boresight effects is given by\,\footnote{For detailed information on this model of the pointing error and its subsequent derivation, one may refer to \cite{yang3}.} \cite[Eq. (5)]{yang3}
\begin{equation}\label{Eq:PE_BS}
\begin{aligned}
f_{p}(I_{p})&=\xi^{2}/A_{0}^{\xi^{2}}\,\exp\left\{-s^{2}/\left(2\,\sigma_s^{2}\right)\right\}\,I_{p}^{\xi^{2}-1}\\
&\times\BesselI{s/\sigma_s\,\sqrt{-2\,\xi^{2}\,\ln\left\{I_{p}/A_{0}\right\}}},\hspace{0.3in} 0\leq I_{p}\leq A_{0},
\end{aligned}
\end{equation}
where $\xi$ is the ratio between the equivalent beam radius at the receiver and the pointing error displacement standard deviation (jitter) $\sigma_{s}$ at the receiver, $A_{0}$ is a constant term that defines the pointing loss, $s$ is the boresight displacement, and $\BesselI[v]{.}$ represents the $v^{\mathrm{th}}$-order modified Bessel functions of an imaginary argument of the first kind \cite[Sec. (8.431)]{gradshteyn}.

\subsubsection{Zero Boresight Pointing Error Model}
The PDF of the irradiance $I_{p}$ with zero boresight effects (i.e. $s=0$ in \eqref{Eq:PE_BS}) is given by\,\footnote{For detailed information on this model of the pointing error and its subsequent derivation, one may refer to \cite{farid}.} \cite[Eq. (11)]{farid}
\begin{equation}\label{Eq:PE}
f_{p}(I_{p})=\xi^{2}\,I_{p}^{\xi^{2}-1}/A_{0}^{\xi^{2}},\hspace{0.3in} 0\leq I_{p}\leq A_{0}.
\end{equation}

\subsection{Atmospheric Turbulence Models}
\subsubsection{Lognormal (LN) Turbulence Scenario}
The optical turbulence can be modeled as LN distribution when the optical channel is considered as a clear-sky atmospheric turbulence channel \cite{niu2}. Hence, for weak turbulence conditions, reference \cite{andrews} suggested a LN PDF to model the irradiance that is the power density of the optical beam. Employing weak turbulence conditions, with a log-scale parameter $\lambda$, the LN PDF of the irradiance $I_{a_{L}}$ is given by (please refer to \cite{andrews,niu2} and references therein)
\begin{equation}\label{Eq:L-N_PDF}
f_{L}(I_{a_{L}})=\frac{1}{I_{a_{L}}\sqrt{2\,\pi}\,\sigma}\,\exp\left\{-\frac{\left[\ln\left\{I_{a_{L}}\right\}-\lambda\right]^{2}}{2\,\sigma^{2}}\right\},\hspace{0.1in}I_{a_{L}}>0,
\end{equation}
where $\sigma^{2}=\mathbb{E}_{I}[I^{2}]/\mathbb{E}_{I}^{2}[I]-1<1$ is defined as the scintillation index \cite[Eq. (1)]{niu2} or the Rytov variance $\sigma_{R}^{2}$ and is related to the log-amplitude variance by $\sigma^{2}_{X}=\sigma_{R}^{2}/4=\sigma^{2}/4$, and $\lambda$ is the log-scale parameter \cite{niu2}.

Now, the joint distribution of $I_{LN}=I_{l}\,I_{a_{L}}\,I_{p}$ can be derived by utilizing
\begin{equation}\label{Eq:Joint_PDF}
\begin{aligned}
f(I_{LN})&=\int_{I_{LN}/A_{0}}^{\infty}f_{I_{LN}|I_{a_{L}}}(I_{LN}|I_{a_{L}})\,f_{L}(I_{a_{L}})\,dI_{a_{L}}\\
&=\int_{I_{LN}/A_{0}}^{\infty}\frac{1}{I_{l}\,I_{a_{L}}}\,f_{p}\left(\frac{I_{LN}}{I_{l}\,I_{a_{L}}}\right)\,f_{L}(I_{a_{L}})\,dI_{a_{L}}.
\end{aligned}
\end{equation}
On substituting \eqref{Eq:L-N_PDF} and \eqref{Eq:PE_BS} appropriately into the integral in \eqref{Eq:Joint_PDF}, following PDF under the influence of nonzero boresight effects is obtained as \cite[Eq. (10)]{yang3}
\begin{equation}\label{Eq:LN_PDF_BS}\small
\begin{aligned}
f(I_{LN})&=\xi^{2}/\left[2\,\left(I_{l}\,A_{0}\right)^{\xi^{2}}\right]\,I_{LN}^{\xi^{2}-1}\,\exp\left\{\xi^{2}\left[\xi^{2}\,\sigma^{2}/2-\lambda\right]+s^{2}/\sigma_{s}^{2}\right\}\\
&\times\mathrm{erfc}\left\{\frac{\xi^{2}\,\sigma^{2}-\lambda+\frac{3\,s}{2\,\xi^{2}\,\sigma_{s}^{2}}+\ln\left\{\frac{I_{LN}}{I_{l}\,A_{0}}\right\}}{\sqrt{2\left(\frac{s^{2}}{\sigma_{s}^{2}\,\xi^{4}}+\sigma^{2}\right)}}\right\},
\end{aligned}
\end{equation}\normalsize
where $\mathrm{erfc}\left\{.\right\}$ is the complementary error function \cite[Eq. (7.1.2)]{abramowitz}. As a special case, for $s=0$, the integral in \eqref{Eq:Joint_PDF} results into the PDF that is in absence of the boresight effects as
\begin{equation}\label{Eq:LN_PDF}
\begin{aligned}
f(I_{LN})&=\xi^{2}/\left[2\,\left(I_{l}\,A_{0}\right)^{\xi^{2}}\right]\,I_{LN}^{\xi^{2}-1}\,\exp\left\{\xi^{2}\left[\xi^{2}\,\sigma^{2}/2-\lambda\right]\right\}\\
&\times\mathrm{erfc}\left\{\left[\xi^{2}\,\sigma^{2}-\lambda+\ln\left\{I_{LN}/\left(I_{l}\,A_{0}\right)\right\}\right]/\left[\sqrt{2}\,\sigma\right]\right\}.
\end{aligned}
\end{equation}

\subsubsection{Rician-Lognormal (RLN) Turbulence Scenario}
In FSO communication environments, the received signals can also be modeled as the product of two independent random processes i.e. a Rician small-scale turbulence process and a lognormal large-scale turbulence process \cite{churnside,yang2}. The Rician PDF (amplitude PDF) of the irradiance $I_{a_{R}}$ is given by \cite[Eq. (2.16)]{slim_book}
\begin{equation}\label{Eq:R_PDF}
\begin{aligned}
f_R\left(I_{a_{R}}\right)&=\left(k^{2}+1\right)/\Omega\,\exp\left\{-k^{2}-\left[\left(k^{2}+1\right)/\Omega\right]\,I_{a_{R}}\right\}\\
&\times\BesselI{2\,k\,\sqrt{\left(k^{2}+1\right)/\Omega\,I_{a_{R}}}},\hspace{0.6in}I_{a_{R}}>0,
\end{aligned}
\end{equation}
where $\Omega$ is the mean-square value or the average power of the irradiance being considered and $0<k<\infty$ is the turbulence parameter. This parameter $k$ is related to the Rician $K$ factor by $K=k^{2}$ that corresponds to the ratio of the power of the line-of-sight (LOS) (specular) component to the average power of the scattered component. The LN PDF is as given in \eqref{Eq:L-N_PDF}.

Now, with the presence of the nonzero boresight pointing errors whose PDF is given in \eqref{Eq:PE_BS}, the combined PDF of $I_{RLN}=I_{l}\,I_{a_{R}}\,I_{a_{L}}\,I_{p}$ is given as
\begin{equation}\label{Eq:RLN_PDF_BS}\tiny
\begin{aligned}
&f\left(I_{RLN}\right)=\left(k^{2}+1\right)\,\xi^{2}/\left[2\,\left(I_{l}\,A_{0}\right)^{\xi^{2}}\right]\,\exp\left\{-k^{2}\right\}\\
&\times\exp\left\{\xi^{2}\left[\frac{\xi^{2}\,\sigma^{2}}{2}-\lambda\right]+\frac{s^{2}}{\sigma_{s}^{2}}\right\}\int_{0}^{\infty}\frac{1}{z^{\,\xi^{2}}}\,\exp\left\{-\frac{k^{2}+1}{z}\,I_{RLN}\right\}\\
&\times\BesselI{2\,k\,\sqrt{\frac{k^{2}+1}{z}\,I_{RLN}}}\mathrm{erfc}\left\{\frac{\xi^{2}\,\sigma^{2}-\lambda+\frac{3\,s}{2\,\xi^{2}\,\sigma_{s}^{2}}+\ln\left\{\frac{z}{I_{l}\,A_{0}}\right\}}{\sqrt{2\left(\frac{s^{2}}{\sigma_{s}^{2}\,\xi^{4}}+\sigma^{2}\right)}}\right\}dz.
\end{aligned}
\end{equation}\normalsize
Similarly, the combined PDF of $I_{RLN}=I_{l}\,I_{a_{R}}\,I_{a_{L}}\,I_{p}$, in presence of zero boresight pointing errors whose PDF is given in \eqref{Eq:PE}, is given as
\begin{equation}\label{Eq:RLN_PDF}\small
\begin{aligned}
&f\left(I_{RLN}\right)=\left(k^{2}+1\right)\,\xi^{2}/\left[2\,\left(I_{l}\,A_{0}\right)^{\xi^{2}}\right]\,\exp\left\{-k^{2}\right\}\\
&\times\exp\left\{\xi^{2}\left[\frac{\xi^{2}\,\sigma^{2}}{2}-\lambda\right]\right\}\int_{0}^{\infty}\frac{1}{z^{\,\xi^{2}}}\,\exp\left\{-\frac{k^{2}+1}{z}\,I_{RLN}\right\}\\
&\times\BesselI{2\,k\,\sqrt{\frac{k^{2}+1}{z}\,I_{RLN}}}\mathrm{erfc}\left\{\frac{\xi^{2}\,\sigma^{2}-\lambda-\ln\left\{\frac{z}{I_{l}\,A_{0}}\right\}}{\sqrt{2}\,\sigma}\right\}dz.
\end{aligned}
\end{equation}\normalsize
The integrals in \eqref{Eq:RLN_PDF_BS} and \eqref{Eq:RLN_PDF}, to the best of our knowledge, are not easy to solve and hence the analysis will be resorted based on moments as will be seen in the upcoming sections.

\subsubsection{M\'{a}laga ($\mathcal{M}$) Turbulence Scenario}
The optical turbulence can be modeled as $\mathcal{M}$ distribution when the irradiance fluctuating of an unbounded optical wavefront (plane or spherical waves) propagates through a turbulent medium under all irradiance conditions in homogeneous, isotropic turbulence \cite{navas}. As a special case, the optical turbulence can be modeled as GG distribution when the optical channel is considered as a cloudy/foggy-sky atmospheric turbulence channel \cite{sandalidis2,tsiftsis,sandalidis,gappmair,wang1}. Hence, employing generalized turbulence conditions, the PDF of the irradiance $I_{a_{M}}$ is given by \cite{navas}
\begin{equation}\label{Eq:M_PDF}\small
f_{M}(I_{a_{M}})=A\sum_{m=1}^{\beta}a_{m}\,I_{a_{M}}\,\BesselK[\alpha-m]{2\sqrt{\frac{\alpha\,\beta\,I_{a_{M}}}{g\,\beta+\Omega^{'}}}},\hspace{0.2in}I_{a_{M}}>0,
\end{equation}\normalsize
where
\begin{equation}
\begin{aligned}
A&\triangleq \frac{2\,\alpha^{\alpha/2}}{g^{1+\alpha/2}\Gamma(\alpha)}\left(\frac{g\,\beta}{g\,\beta+\Omega^{'}}\right)^{\beta+\alpha/2},\\
a_{m}&\triangleq \binom{\beta-1}{m-1}\frac{\left(g\,\beta+\Omega^{'}\right)^{1-m/2}}{\left(m-1\right)!}\left(\frac{\Omega^{'}}{g}\right)^{m-1}\left(\frac{\alpha}{\beta}\right)^{m/2},
\end{aligned}
\end{equation}
$\alpha$ is a positive parameter related to the effective number of large-scale cells of the scattering process, $\beta$ is the amount of fading parameter and is a natural number\,\footnote{A generalized expression of \eqref{Eq:M_PDF} is given in \cite[Eq. (22)]{navas} for $\beta$ being a real number though it is less interesting due to the high degree of freedom of the proposed distribution (Sec. III of \cite{navas}).}, $g=\Expected{|U_{S}^{G}|^{2}}=2\,b_{0}\,(1-\rho)$ denotes the average power of the scattering component received by off-axis eddies, $2\,b_{0}=\Expected{|U_{S}^{C}|^{2}+|U_{S}^{G}|^{2}}$ is the average power of the total scatter components, the parameter $0\leq\rho\leq 1$ represents the amount of scattering power coupled to the LOS component, $\Omega^{'}=\Omega+2\,b_{0}\,\rho+2\sqrt{2\,b_{0}\,\rho\,\Omega}\cos(\phi_{A}-\phi_{B})$ represents the average power from the coherent contributions, $\Omega=\Expected{|U_{L}|^{2}}$ is the average power of the LOS component, $\phi_{A}$ and $\phi_{B}$ are the deterministic phases of the LOS and the coupled-to-LOS scatter terms, respectively, $\Gamma(.)$ is the Gamma function as defined in \cite[Eq. (8.310)]{gradshteyn}, and $K_{v}(.)$ is the $v^{\mathrm{th}}$-order modified Bessel function of the second kind \cite[Sec. (8.432)]{gradshteyn}. It is interesting to know here that $\Expected{|U_{S}^{C}|^{2}}=2\,b_{0}\,\rho$ denotes the average power of the coupled-to-LOS scattering component and $\Expected{I_{a}}=\Omega+2\,b_{0}$.\footnote{Detailed information on the $\mathcal{M}$ distribution, its formation, and its random generation can be extracted from \cite[Eqs. (13-21)]{navas}.}

Now, with the presence of the nonzero boresight pointing errors whose PDF is given in \eqref{Eq:PE_BS}, the combined PDF of $I_{M}=I_{l}\,I_{a_{M}}\,I_{p}$ is given as
\begin{equation}\label{Eq:M_PDF_BS}\footnotesize
\begin{aligned}
&f\left(I_{M}\right)=\frac{\xi^{2}\,A\,I_{M}^{\xi^{2}-1}}{I_{l}^{\xi^{2}}A_{0}^{\xi^{2}}}\,\exp\left\{-\frac{s^{2}}{2\,\sigma_s^{2}}\right\}\sum_{m=1}^{\beta}\int_{I/A_{0}}^{\infty}I_{a{M}}^{1-\xi^{2}}\\
&\times\BesselI{\frac{s}{\sigma_s}\,\sqrt{-2\,\xi^{2}\,\ln\left\{\frac{I_{M}}{I_{l}\,I_{a_{M}}\,A_{0}}\right\}}}\,\BesselK[\alpha-m]{2\sqrt{\frac{\alpha\,\beta\,I_{a_{M}}}{g\,\beta+\Omega^{'}}}}dI_{a_{M}}.
\end{aligned}
\end{equation}\normalsize
The integral in \eqref{Eq:M_PDF_BS}, to the best of our knowledge, is not easy to solve in closed-form and hence the analysis will be resorted based on moments as will be seen in the upcoming sections. Similarly, the combined PDF of $I_{M}=I_{l}\,I_{a_{M}}\,I_{p}$, in presence of zero boresight pointing errors (i.e. $s=0$ in \eqref{Eq:M_PDF_BS}) whose PDF is given in \eqref{Eq:PE}, is known to be given by \cite{navas}
\begin{equation}\label{Eq:M_PDF}
f(I_{M})=\frac{\xi^{2}A}{2\,I_{M}}\sum_{m=1}^{\beta}b_{m}\,\MeijerG[right]{3,0}{1,3}{\frac{\alpha\,\beta}{\left(g\,\beta+\Omega^{'}\right)}\frac{I_{M}}{A_{0}}}{\xi^2+1}{\xi^2,\alpha,m},
\end{equation}
where $b_{m}=a_{m}\left[\alpha\,\beta/\left(g\,\beta+\Omega^{'}\right)\right]^{-\left(\alpha+m\right)/2}$ and $\mathrm{G}[.]$ is the Meijer's G function as defined in \cite[Eq. (9.301)]{gradshteyn}.

\subsection{Important Outcomes and Further Motivations}
To the best of our knowledge, it is quite tedious to utilize these expressions in \eqref{Eq:LN_PDF_BS}, \eqref{Eq:LN_PDF}, \eqref{Eq:RLN_PDF_BS}, \eqref{Eq:RLN_PDF}, \eqref{Eq:M_PDF_BS}, and \eqref{Eq:M_PDF}\,\footnote{Similar results corresponding to \eqref{Eq:M_PDF_BS} and \eqref{Eq:M_PDF} have also been derived for the GG turbulence scenario though those have not been presented here as GG turbulence is a special case of $\mathcal{M}$ turbulence.}. As will be shown in Section IV, it is in most cases not possible or challenging to deal with such expressions to obtain some further exact closed-form results for the ergodic capacity of such a FSO channel. Therefore, the capacity analysis of such FSO link, in a simpler way, is carried out utilizing moments as will be derived in the following section.

\section{Exact Closed-Form Moments}
As we have seen above that it is quite a challenge to obtain closed-form PDF and even if we are able to find one, the expression(s) are not simple enough to be utilized further for the analysis of the ergodic capacity as will be seen in the following section. Hence, we resort to moments based analysis for which the moments for the various turbulence scenarios discussed in the previous section are derived here.

For the heterodyne detection technique case, the instantaneous SNR $\gamma=\eta_{e}\,I/N_{0}$ and the average SNR\,\footnote{$\overline{\gamma}_{\mathrm{heterodyne}}$ is the average SNR for coherent/heterodyne FSO systems given by $\overline{\gamma}_{\mathrm{heterodyne}}=C_{c}$ \cite[Eq. (7)]{niu1}, where $C_{c}=2\,R^{2}A\,P_{LO}/\left[2\,q\,R\,\Delta f\,P_{LO}+2\,\Delta f\left(q\,R\,A\,I_{b}+2\,k_{b}\,T_{k}\,F_{n}/R_{L}\right)\right]\approx R\,A/\left(q\,\Delta f\right)$ is a multiplicative constant for a given heterodyne/coherent system, where $R$ is the photodetector responsivity, $A$ is the photodetector area, $P_{LO}$ is the local oscillator power, $\Delta f$ denotes the noise equivalent bandwidth of a FSO receiver, $q$ is the electronic charge, $I_{b}$ is the background light irradiance, $k_{b}$ is Boltzmann's constant, $T_{k}$ is the temperature in Kelvin, $F_{n}$ represents a thermal noise enhancement factor due to amplifier noise, and $R_{L}$ is the load resistance. It is evident that $C_{c}=\mu_{\mathrm{heterodyne}}$ in this work.} develops as $\mu_{\mathrm{heterdoyne}}=\mathbb{E}_{\gamma_{\mathrm{heterodyne}}}[\gamma]=\overline{\gamma}_{\mathrm{heterodyne}}=\eta_{e}\,\mathbb{E}_{I}[I]/N_{0}$, where $\eta_{e}$ is the effective photoelectric conversion ratio, $N_{0}$ symbolizes the AWGN sample, and $\Expected{.}$ denotes the expectation operator.

Similarly, for the IM/DD technique, $\gamma=\eta_{e}^{2}\,I^{2}/N_{0}$ and the electrical SNR\,\footnote{$\overline{\gamma}_{\mathrm{IM/DD}}$ is the average SNR for IM/DD FSO systems given by $\overline{\gamma}_{\mathrm{IM/DD}}=C_{s}\,\mathbb{E}_{I}[I^{2}]/\mathbb{E}_{I}^{2}[I]$, where $C_{s}=\left(R\,A\,\xi\right)^{2}/\left[2\,\Delta f\left(q\,R\,A\,I_{b}+2\,k_{b}\,T_{k}\,F_{n}/R_{L}\right)\right]$ \cite{niu1} is a multiplicative constant for a given IM/DD system. It is evident that $C_{s}=\mu_{\mathrm{IM/DD}}$ in this work.} develops as $\mu_{\mathrm{IM/DD}}=\mathbb{E}_{\gamma_{\mathrm{IM/DD}}}[\gamma]\,\mathbb{E}_{I}^{2}[I]/\mathbb{E}_{I}[I^{2}]=\overline{\gamma}_{\mathrm{IM/DD}}\,\mathbb{E}_{I}^{2}[I]/\mathbb{E}_{I}[I^{2}]=\eta_{e}^{2}\,\mathbb{E}_{I}^{2}[I]/N_{0}$ \cite{gappmair}.

Now, on unifying the SNR expressions above for both the detection types, $\gamma_{r}=\eta_{e}^{r}\,I^{r}/N_{0}$ and $\mu_{r}=\eta_{e}^{r}\,\mathbb{E}_{I}^{r}[I]/N_{0}$ are obtained. Since, $I_{a}$ and $I_{p}$ are independent random variables, the unified moments are defined as\,\footnote{$I_{l}$, $A_{0}$, and $\lambda$ cancel out being deterministic parameters.}$^{,}$\,\footnote{$\overline{\gamma}_{1}$ is the first moment (i.e. $n=1$) for the heterodyne ($r=1$) case as can be seen from \eqref{Eq:Moments_Def}. Based on this substitution, we obtain $\overline{\gamma}_{1}=\mu_{1}$ signifying that $\overline{\gamma}_{1}$ and $\mu_{1}$ are the same quantity defined as the average SNR for the heterodyne FSO systems. Similarly, $\overline{\gamma}_{2}$ is the first moment (i.e. $n=1$) for the IM/DD ($r=2$) case as can be seen from \eqref{Eq:Moments_Def}. Based on this substitution, we obtain $\overline{\gamma}_{2}=\Expected{I_{a}^{2}}\,\Expected{I_{p}^{2}}/\left(\mathbb{E}^{2}[I_{a}]\,\mathbb{E}^{2}[I_{p}]\right)\,\mu_{2}=\Expected{I^{2}}/\mathbb{E}^{2}[I]\,\mu_{2}$ or $\mu_{2}=\mathbb{E}^{2}[I_{a}]\,\mathbb{E}^{2}[I_{p}]/\left(\Expected{I_{a}^{2}}\,\Expected{I_{p}^{2}}\right)\,\overline{\gamma}_{2}=\mathbb{E}^{2}[I]/\Expected{I^{2}}\,\overline{\gamma}_{2}$ signifying that $\overline{\gamma}_{2}$ and $\mu_{2}$ are different quantities defined as the average SNR and the electrical SNR for the IM/DD FSO systems, respectively \cite{niu1}.}
\begin{equation}\label{Eq:Moments_Def}
\begin{aligned}
\Expected{\gamma^n_{r}}&=\eta_{e}^{r\,n}\,\Expected{I^{r\,n}}/N_{0}^{n}=\mu_{r}^{n}\,\Expected{\left(I_{a}\,I_{p}\right)^{r\,n}}/\mathbb{E}^{r\,n}[I_{a}\,I_{p}]\\
&=\mu_{r}^{n}\,\Expected{I_{a}^{r\,n}}\,\Expected{I_{p}^{r\,n}}/\left(\mathbb{E}^{r\,n}[I_{a}]\,\mathbb{E}^{r\,n}[I_{p}]\right).
\end{aligned}
\end{equation}

\subsection{Lognormal (LN) Turbulence Scenario}
The unified moments for this particular scenario are defined as
\begin{equation}\label{Eq:Moments_Def_LN}\small
\begin{aligned}
\Expected{\gamma^n_{r}}_{LN}&=\eta_{e}^{r\,n}\,\Expected{I^{r\,n}}/N_{0}^{n}=\mu_{r}^{n}\,\Expected{\left(I_{a_{L}}\,I_{p}\right)^{r\,n}}/\mathbb{E}^{r\,n}[I_{a_{L}}\,I_{p}]\\
&=\mu_{r}^{n}\,\Expected{I_{a_{L}}^{r\,n}}\,\Expected{I_{p}^{r\,n}}/\left(\mathbb{E}^{r\,n}[I_{a_{L}}]\,\mathbb{E}^{r\,n}[I_{p}]\right).
\end{aligned}
\end{equation}\normalsize
Utilizing the definition of the moments, $\Expected{I_{a_{L}}^{r\,n}}$ and $\Expected{I_{p}^{r\,n}}$ for nonzero boresight pointing errors are easily obtained after some manipulations as $\Expected{I_{a_{L}}^{r\,n}}=\exp\left\{r\,n\,\lambda+\left(r\,n\,\sigma\right)^{2}/2\right\}$ and $\Expected{I_{p}^{r\,n}}=A_{0}^{r\,n}\,\xi^{2}/\left(\xi^{2}+r\,n\right)\,\exp\left\{-r\,n\,s^{2}/\left[2\,\sigma_s^{2}\left(\xi^{2}+r\,n\right)\right]\right\}$ \cite[Eq. (6)]{yang3}, respectively. Substituting these back into \eqref{Eq:Moments_Def_LN}, the unified exact closed-form moments for LN atmospheric turbulence in presence of nonzero boresight pointing errors are obtained as
\begin{equation}\label{Eq:Moments_LN_BS}
\begin{aligned}
&\Expected{\gamma^n_{r}}_{LN}=\frac{\xi^{2(1-r\,n)}}{\left(\xi^{2}+r\,n\right)\left(\xi^{2}+1\right)^{-r\,n}}\,\exp\left\{\frac{r\,n\,\sigma^{2}}{2}\left(r\,n-1\right)\right.\\
&+\left.r\,n\,s^{2}/\left(2\,\sigma_s^{2}\right)\left[1/\left(\xi^{2}+1\right)-1/\left(\xi^{2}+r\,n\right)\right]\right\}\,\mu_{r}^{n}.
\end{aligned}
\end{equation}
Similarly, when considering zero boresight pointing errors (i.e. special case with $s=0$), the $\Expected{I_{p}^{r\,n}}=A_{0}^{r\,n}\,\xi^{2}/\left(\xi^{2}+r\,n\right)$ and the corresponding unified exact closed-form moments for LN atmospheric turbulence in presence of zero boresight pointing errors are obtained as
\begin{equation}\label{Eq:Moments_LN}\small
\Expected{\gamma^n_{r}}_{LN}=\frac{\xi^{2(1-r\,n)}}{\left(\xi^{2}+r\,n\right)\left(\xi^{2}+1\right)^{-r\,n}}\,\exp\left\{\frac{r\,n\,\sigma^{2}}{2}\left(r\,n-1\right)\right\}\,\mu_{r}^{n}.
\end{equation}\normalsize

\subsection{Rician-Lognormal (RLN) Turbulence Scenario}
Since $I_{a_{R}}$, $I_{a_{L}}$, and $I_{p}$ are independent random variables, the unified moments for RLN turbulence scenario are defined as
\begin{equation}\label{Eq:Moments_Def_RLN}\small
\begin{aligned}
&\Expected{\gamma^n_{r}}_{RLN}=\eta_{e}^{r\,n}\,\Expected{I^{r\,n}}/N_{0}^{n}\\
&=\mu_{r}^{n}\,\Expected{\left(I_{a_{R}}\,I_{a_{L}}\,I_{p}\right)^{r\,n}}/\mathbb{E}^{r\,n}[I_{a_{R}}\,I_{a_{L}}\,I_{p}]\\
&=\mu_{r}^{n}\,\Expected{I_{a_{R}}^{r\,n}}\,\Expected{I_{a_{L}}^{r\,n}}\,\Expected{I_{p}^{r\,n}}/\left(\mathbb{E}^{r\,n}[I_{a_{R}}]\,\mathbb{E}^{r\,n}[I_{a_{L}}]\,\mathbb{E}^{r\,n}[I_{p}]\right).
\end{aligned}
\end{equation}\normalsize
Utilizing the definition of the moments, $\Expected{I_{a_{L}}^{r\,n}}$ and $\Expected{I_{p}^{r\,n}}$ for nonzero boresight pointing errors were easily obtained in previous subsection i.e. Section III.A whereas
$\Expected{I_{a_{R}}^{r\,n}}=\left[\Omega/\left(k^{2}+1\right)\right]^{r\,n}\,\Gamma\left(r\,n+1\right)\,\Hypergeom{1}{1}{-r\,n}{1}{-k^{2}}$ \cite[Eq. (2.18)]{slim_book}, where $\Hypergeom{p}{q}{.}{.}{.}$ represents the generalized hypergeometric $F$ function \cite[Eq. (9.14.1)]{gradshteyn} and more specifically, $\Hypergeom{1}{1}{.}{.}{.}$ represents the confluent hypergeometric $F$ function \cite[Eq. (9.210.1)]{gradshteyn}. Substituting these back into \eqref{Eq:Moments_Def_RLN}, the unified exact closed-form moments for RLN turbulence under nonzero boresight pointing errors are obtained as\,\footnote{It must be noted that $\Hypergeom{1}{1}{-1}{1}{-k^{2}}=k^{2}+1$.}
\begin{equation}\label{Eq:Moments_RLN_BS}
\begin{aligned}
&\Expected{\gamma^n_{r}}_{RLN}=\xi^{2(1-r\,n)}/\left[\left(\xi^{2}+r\,n\right)\left(\xi^{2}+1\right)^{-r\,n}\right]\\
&\times\exp\left\{\frac{r\,n\,\sigma^{2}}{2}\left(r\,n-1\right)+\frac{r\,n\,s^{2}}{2\,\sigma_s^{2}}\left(\frac{1}{\xi^{2}+1}-\frac{1}{\xi^{2}+r\,n}\right)\right\}\\
&\times\Gamma\left(r\,n+1\right)\,\Hypergeom{1}{1}{-r\,n}{1}{-k^{2}}/\left(k^{2}+1\right)^{r\,n}\,\mu_{r}^{n}.
\end{aligned}
\end{equation}
Similarly, when considering zero boresight pointing errors (i.e. special case with $s=0$), the corresponding unified exact closed-form moments for RLN atmospheric turbulence in presence of zero boresight pointing errors are obtained as
\begin{equation}\label{Eq:Moments_RLN}
\begin{aligned}
&\Expected{\gamma^n_{r}}_{RLN}=\xi^{2(1-r\,n)}/\left[\left(\xi^{2}+r\,n\right)\left(\xi^{2}+1\right)^{-r\,n}\right]\\
&\times\exp\left\{\frac{r\,n\,\sigma^{2}}{2}\left(r\,n-1\right)\right\}\frac{\Hypergeom{1}{1}{-r\,n}{1}{-k^{2}}}{\left(k^{2}+1\right)^{r\,n}\,\Gamma\left(r\,n+1\right)^{-1}}\,\mu_{r}^{n}.
\end{aligned}
\end{equation}

\subsection{M\'{a}laga ($\mathcal{M}$) Turbulence Scenario}
Since $I_{a_{M}}$ and $I_{p}$ are independent random variables, the unified moments for $\mathcal{M}$ turbulence scenario are defined as
\begin{equation}\label{Eq:Moments_Def_M}\small
\begin{aligned}
\Expected{\gamma^n_{r}}_{M}&=\eta_{e}^{r\,n}\,\Expected{I^{r\,n}}/N_{0}^{n}=\mu_{r}^{n}\,\Expected{\left(I_{a_{M}}\,I_{p}\right)^{r\,n}}/\mathbb{E}^{r\,n}[I_{a_{M}}\,I_{p}]\\
&=\mu_{r}^{n}\,\Expected{I_{a_{M}}^{r\,n}}\,\Expected{I_{p}^{r\,n}}/\left(\mathbb{E}^{r\,n}[I_{a_{M}}]\,\mathbb{E}^{r\,n}[I_{p}]\right).
\end{aligned}
\end{equation}\normalsize
Utilizing the definition of the moments, $\Expected{I_{p}^{r\,n}}$ for nonzero boresight pointing errors was easily obtained in previous subsection i.e. Section III.A whereas $\Expected{I_{a_{M}}^{r\,n}}/\mathbb{E}^{r\,n}[I_{a_{M}}]=r\,A\,\Gamma(r\,n+\alpha)\sum_{m=1}^{\beta}\,b_{m}\,\Gamma(r\,n+m)/\left(2^{r}\,B^{r\,n}\right)$ where $B=\alpha\,\beta\,(g+\Omega^{'})/(g\,\beta+\Omega^{'})$. Substituting these back into \eqref{Eq:Moments_Def_M}, the unified exact closed-form moments for $\mathcal{M}$ turbulence under nonzero boresight pointing errors are obtained as
\begin{equation}\label{Eq:Moments_M_BS}
\begin{aligned}
&\Expected{\gamma^n_{r}}_{M}=\xi^{2(1-r\,n)}/\left[\left(\xi^{2}+r\,n\right)\left(\xi^{2}+1\right)^{-r\,n}\right]\\
&\times\,\exp\left\{r\,n\,s^{2}/\left(2\,\sigma_s^{2}\right)\left[1/\left(\xi^{2}+1\right)-1/\left(\xi^{2}+r\,n\right)\right]\right\}\\
&\times r\,A\,\Gamma(r\,n+\alpha)/\left(2^{r}\,B^{r\,n}\right)\sum_{m=1}^{\beta}\,b_{m}\,\Gamma(r\,n+m)\,\mu_{r}^{n}.
\end{aligned}
\end{equation}
As a special case, the unified exact closed-form moments for GG turbulence under nonzero boresight pointing errors are obtained as
\begin{equation}\label{Eq:Moments_G_BS}
\begin{aligned}
&\Expected{\gamma^n_{r}}_{GG}=\frac{\xi^{2\left(1-r\,n\right)}\,\left(\xi^{2}+1\right)^{r\,n}\,\Gamma\left(r\,n+\alpha\right)\,\Gamma\left(r\,n+\beta\right)}{\left(\xi^{2}+r\,n\right)\,\left(\alpha\,\beta\right)^{r\,n}\,\Gamma\left(\alpha\right)\,\Gamma\left(\beta\right)}\\
&\times\,\exp\left\{r\,n\,s^{2}/\left(2\,\sigma_s^{2}\right)\left[1/\left(\xi^{2}+1\right)-1/\left(\xi^{2}+r\,n\right)\right]\right\}\,\mu_{r}^{n}.
\end{aligned}
\end{equation}
Similarly, when considering zero boresight pointing errors (i.e. special case with $s=0$), the corresponding unified exact closed-form moments for $\mathcal{M}$ atmospheric turbulence in presence of zero boresight pointing errors are obtained as
\begin{equation}\label{Eq:Moments_M}
\Expected{\gamma^n_{r}}_{M}=\frac{r\,\xi^2\,A\,\Gamma(r\,n+\alpha)}{2^{r}\,\left(r\,n+\xi^2\right)\,B^{r\,n}}\sum_{m=1}^{\beta}b_{m}\,\Gamma(r\,n+m)\,\mu_{r}^{\,n}.
\end{equation}
As a special case, the corresponding unified exact closed-form moments for GG atmospheric turbulence in presence of zero boresight pointing errors are obtained as
\begin{equation}\label{Eq:Moments_G}
\Expected{\gamma^n_{r}}_{GG}=\frac{\xi^{2\left(1-r\,n\right)}\,\left(\xi^{2}+1\right)^{r\,n}\,\Gamma(r\,n+\alpha)\Gamma(r\,n+\beta)}{\left(\xi^2+r\,n\right)\,\left(\alpha\,\beta\right)^{r\,n}\Gamma(\alpha)\,\Gamma(\beta)}\,\mu_{r}^{\,n}.
\end{equation}

\subsection{Important Outcomes and Further Motivations}
\begin{itemize}
\item Interestingly enough and expectedly, these expressions in \eqref{Eq:Moments_LN_BS}, \eqref{Eq:Moments_LN}, \eqref{Eq:Moments_RLN_BS}, \eqref{Eq:Moments_RLN}, and \eqref{Eq:Moments_M_BS}-\eqref{Eq:Moments_G} reduce to only $\mu_{1}^{n}$ for $r=1$ (heterodyne detection technique) case thereby supporting the difference between the definitions of average SNR vs. electrical SNR.
\item It is worthy to note that these simple results for the moments can be directly plugged into \cite[Eq. (3)]{yilmaz3} to obtain the $n^{\mathrm{th}}$-order amount of fading for the instantaneous SNR, $\gamma$. These interesting results can be then utilized to parameterize the distribution of the SNR of the received signal.
\item More importantly, these simple results for the moments are useful to conduct asymptotic analysis of the ergodic capacity as shown in the following section of this work.
\end{itemize}

\section{Ergodic Capacity}
\subsection{General Methodology}
The ergodic channel capacity $\overline{C}$ is defined as \cite[Eq. (26)]{lapidoth},\cite[Eq. (7.43)]{owc}
\begin{equation}\label{Eq:EC}
\overline{C} \triangleq \Expected{\ln\left\{1+c\,\gamma\right\}},
\end{equation}
where $c$ is a constant term such that $c=1$ for heterodyne detection giving an exact result and $c=e/\left(2\,\pi\right)$ for IM/DD giving a lower-bound result \cite{lapidoth,owc}\,\footnote{For readers clarification, to the best of the authors' knowledge based on the open literature, there does not exists any actual mathematical formulation for analyzing the ergodic capacity of such FSO channels.}. Additionally, knowing that $I_{a}$ and $I_{p}$ are independent random variables, we can re-write the definition of the ergodic capacity as
\begin{equation}\label{Eq:EC_Int}\small
\begin{aligned}
\overline{C}&=\Expected{\ln\left\{1+\frac{c\,\left(\eta_{e}\,I\right)^{r}}{N_{0}}\right\}}=\int_{0}^{\infty}\,\ln\left\{1+\frac{c\,\left(\eta_{e}\,I\right)^{r}}{N_{0}}\right\}\,f\left(I\right)\,dI\\
&=\int_{0}^{\infty}\int_{0}^{A_{0}}\,\ln\left\{1+\frac{c\,\left(\eta_{e}\,I_{l}\,I_{a}\,I_{p}\right)^{r}}{N_{0}}\right\}\,f_a\left(I_{a}\right)\,f_p\left(I_{p}\right)\,dI_{p}\,dI_{a}.
\end{aligned}
\end{equation}\normalsize
Since, $I_{p}$ is the common random variable in all the different atmospheric turbulence scenarios, we can try to solve \eqref{Eq:EC_Int} for the two types of pointing errors. By placing \eqref{Eq:PE_BS} into \eqref{Eq:EC_Int}, to the best of our knowledge, it is not possible to find an exact closed-form solution for the inner integral. On the other hand, if we place \eqref{Eq:PE} into \eqref{Eq:EC_Int}, we obtain
\begin{equation}\label{Eq:EC_Int2}
\begin{aligned}
\overline{C}&=\int_{0}^{\infty}\int_{0}^{A_{0}}\,\ln\left\{1+\frac{c\,\left(\eta_{e}\,I_{l}\,I_{a}\,I_{p}\right)^{r}}{N_{0}}\right\}\,\xi^{2}\,I_{p}^{\xi^{2}-1}/A_{0}^{\xi^{2}}\,dI_{p}\\
&\times f_a\left(I_{a}\right)\,dI_{a}=\int_{0}^{\infty}\,\left[\ln\left\{\frac{c\,\left(\eta_{e}\,A_{0}\,I_{l}\,I_{a}\right)^{r}}{N_{0}}+1\right\}\right.\\
&\left.-\frac{c\,\left(\eta_{e}\,A_{0}\,I_{l}\,I_{a}\right)^{r}}{N_{0}}\,\Phi\left(-\frac{c\,\left(\eta_{e}\,A_{0}\,I_{l}\,I_{a}\right)^{r}}{N_{0}},1,\frac{\xi^{2}+r}{r}\right)\right]\\
&\times f_a\left(I_{a}\right)\,dI_{a},
\end{aligned}
\end{equation}
where $\Phi\left(.\right)$ is the LerchPhi function \cite[Eq. (10.06.02.0001.01)]{mathematica}.

If an exact closed-form is not obtainable via either \eqref{Eq:EC} and/or \eqref{Eq:EC_Int} and/or \eqref{Eq:EC_Int2}, the ergodic capacity can be analyzed utilizing the moments. At \textit{\textbf{high SNR}}, an asymptotic analysis can be done by utilizing the moments yielding an asymptotically tight lower bound given by\,\footnote{For readers clarification, it is possible to use SNR moments as an efficient tool for deriving even higher order ergodic capacity statistics via utilizing \cite[Eq. (6)]{yilmaz3}.} \cite[Eqs. (8) and (9)]{yilmaz3}
\begin{equation}\label{Eq:EC_Single_High_Def}
\overline{C}\underset{\mu_{r}\,>>1}{\approxeq}\ln(c\,\mu_{r})+\zeta,
\end{equation}
where
\begin{equation}\label{Eq:AF_Single_Derivative_Def}
\zeta=\left.\partial/\partial n\left(\Expected{\gamma_{r}^n}/\Expected{\gamma_{r}}^n-1\right)\right|_{n=0}.
\end{equation}
The expression in \eqref{Eq:EC_Single_High_Def} can be simplified to
\begin{equation}\label{Eq:EC_Single_High_Def_Simplify}\small
\overline{C}\underset{\mu_{r}\,>>1}{\approxeq}\ln(c\,\mu_{r})+\left.\frac{\partial}{\partial n}\left(\frac{\Expected{\gamma_{r}^n}}{\Expected{\gamma_{r}}^n}-1\right)\right|_{n=0}=\left.\frac{\partial}{\partial n}\,\Expected{\gamma_{r}^n}\right|_{n=0}.
\end{equation}\normalsize
Similarly, at \textit{\textbf{low SNR}}, it can be easily shown that the ergodic capacity can be asymptotically approximated by the first moment in closed-form.

\subsection{Lognormal (LN) Turbulence Scenario}
\subsubsection{Exact Analysis}
For LN atmospheric turbulence scenario under nonzero boresight pointing errors and zero boresight pointing errors, we respectively substitute \eqref{Eq:LN_PDF_BS} and \eqref{Eq:LN_PDF} in \eqref{Eq:EC}. To the best of our knowledge, both the above scenarios can not be solved in exact closed-form.

Additionally, we have already reached to a conclusion that it is not possible to solve the inner integral for nonzero boresight pointing errors in \eqref{Eq:EC_Int} with \eqref{Eq:PE_BS}. Alternatively, by substituting \eqref{Eq:L-N_PDF} in \eqref{Eq:EC_Int}, the outer integral for LN PDF $f_{L}\left(I_{a_{L}}\right)$ in \eqref{Eq:EC_Int} does not lead to possible exact closed-form results. On the other hand, we have been able to solve the inner integral for zero boresight pointing errors in \eqref{Eq:EC_Int} with \eqref{Eq:PE} to obtain \eqref{Eq:EC_Int2} and hence on placing the LN PDF $f_{L}\left(I_{a_{L}}\right)$ \eqref{Eq:L-N_PDF} into \eqref{Eq:EC_Int2}, we obtain
\begin{equation}\label{Eq:EC_Int2_LN}
\begin{aligned}
\overline{C}&=\frac{1}{\sqrt{2\,\pi}\,\sigma}\int_{0}^{\infty}\frac{1}{I_{a_{L}}}\,\exp\left\{-\left[\frac{\ln\left\{I_{a_{L}}\right\}-\lambda}{\sqrt{2}\,\sigma}\right]^{2}\right\}\\
&\times\left[\ln\left\{\frac{c\,\left(\eta_{e}\,A_{0}\,I_{l}\,I_{a_{L}}\right)^{r}}{N_{0}}+1\right\}-\frac{c\,\left(\eta_{e}\,A_{0}\,I_{l}\,I_{a_{L}}\right)^{r}}{N_{0}}\right.\\
&\times\left.\Phi\left(-\frac{c\,\left(\eta_{e}\,A_{0}\,I_{l}\,I_{a_{L}}\right)^{r}}{N_{0}},1,\frac{\xi^{2}+r}{r}\right)\right]\,dI_{a_{L}}.
\end{aligned}
\end{equation}
On applying simple change of random variable $x=\left(\ln\left\{I_{a_{L}}\right\}-\lambda\right)/\left(\sqrt{2}\,\sigma\right)$, we get $I_{a_{L}}=\exp\left\{\sqrt{2}\,\sigma\,x+\lambda\right\}$ and $dI_{a_{L}}=\sqrt{2}\,\sigma\,\exp\left\{\sqrt{2}\,\sigma\,x+\lambda\right\}\,dx$ leading to
\begin{equation}\label{Eq:EC_Int2_LN_GH}
\overline{C}=\frac{1}{\sqrt{\pi}}\int_{-\infty}^{\infty}\exp\left\{-x^{2}\right\}\,f_{x}\left(x\right)\,dx,
\end{equation}
where
\begin{equation}\label{Eq:fx}
\begin{aligned}
&f_{x}\left(x\right)=\ln\left\{\frac{c\,\left(\eta_{e}\,A_{0}\,I_{l}\right)^{r}}{N_{0}}\,\exp\left\{r\left(\sqrt{2}\,\sigma\,x+\lambda\right)\right\}+1\right\}\\
&-\frac{c\,\left(\eta_{e}\,A_{0}\,I_{l}\right)^{r}}{N_{0}}\,\exp\left\{r\left(\sqrt{2}\,\sigma\,x+\lambda\right)\right\}\\
&\times\Phi\left(-\frac{c\,\left(\eta_{e}\,A_{0}\,I_{l}\right)^{r}}{N_{0}}\,\exp\left\{r\left(\sqrt{2}\,\sigma\,x+\lambda\right)\right\},1,\frac{\xi^{2}+r}{r}\right).
\end{aligned}
\end{equation}
The integral in \eqref{Eq:EC_Int2_LN_GH} is solvable with the help of $N=20$-point Gauss-Hermite formula \cite[Eq. (25.4.46)]{abramowitz} leading to
\begin{equation}\label{Eq:EC_Int2_LN_GH2}
\overline{C}\approxeq\frac{1}{\sqrt{\pi}}\,\sum_{i=1}^{N}\,w_{i}\,f_{x}\left(x_{i}\right),
\end{equation}
where $w_{i}$ and $x_{i}$ are the weights and the abscissas that can be acquired from \cite[Table 25.10]{abramowitz}.

\subsubsection{Approximate Analysis}
Reverting back to LN atmospheric turbulence under nonzero boresight pointing errors, since it is not feasible to obtain an exact closed-form solution, we utilize the moments derived earlier to deduce the asymptotic results. Hence, based on \eqref{Eq:EC_Single_High_Def_Simplify}, the first derivative of the moments in \eqref{Eq:Moments_LN_BS} is required to be evaluated at $n=0$ for high SNR asymptotic approximation to the ergodic capacity. The first derivative of the moments in \eqref{Eq:Moments_LN_BS} is given as
\begin{equation}\label{Eq:Moments_Derivative_LN_BS}\small
\begin{aligned}
&\partial/\partial n\,\Expected{\gamma_{r}^n}=\xi^{2(1-r\,n)}/\left[\left(\xi^{2}+r\,n\right)\left(\xi^{2}+1\right)^{-r\,n}\right]\\
&\times\exp\left\{\frac{r\,n\,\sigma^{2}}{2}\left(r\,n-1\right)+\frac{r\,n\,s^{2}}{2\,\sigma_s^{2}}\left(\frac{1}{\xi^{2}+1}-\frac{1}{\xi^{2}+r\,n}\right)\right\}\\
&\times\left\{r\,\sigma^{2}\left(\,r\,n-\frac{1}{2}\right)+\frac{r\,s^{2}}{2\,\sigma_s^{2}}\left[\frac{r\,n}{\left(\xi^{2}+r\,n\right)^{2}}+\frac{1}{\xi^{2}+1}-\frac{1}{\xi^{2}+r\,n}\right]\right.\\
&-\left.r/\left(r\,n+\xi^{2}\right)-r\,\ln\left\{\xi^{2}/\left(\xi^{2}+1\right)\right\}+\ln\left\{c\,\mu_{r}\right\}\right\}\left(c\,\mu_{r}\right)^{n},
\end{aligned}
\end{equation}\normalsize
and at $n=0$, it evaluates to
\begin{equation}\label{Eq:Moments_Derivative_n0_LN_BS}
\begin{aligned}
\overline{C}\underset{\mu_{r}\,>>1}{\approxeq}&\ln\left\{c\,\mu_{r}\right\}-r\left[\frac{1}{\xi^{2}}+\frac{\sigma^2}{2}+\frac{s^{2}}{2\,\sigma_s^{2}\,\xi^{2}\,\left(\xi^{2}+1\right)}\right.\\
&\left.+\ln\left\{\xi^2/\left(\xi^2+1\right)\right\}\right].
\end{aligned}
\end{equation}
Similarly, for LN atmospheric turbulence under zero boresight pointing errors (i.e. for $s=0$) and under no pointing errors (i.e. for $s=0$ and $\xi\rightarrow\infty$), the asymptotic approximations to their respective ergodic capacity's at high SNR are derived in \tableref{Tab:Special_Cases}.

Furthermore, for \textit{\textbf{low SNR}} asymptotic analysis, it can be easily shown that the ergodic capacity can be asymptotically approximated by the first moment. Utilizing \eqref{Eq:Moments_LN_BS} via placing $n=1$ in it, the ergodic capacity of a single FSO link under LN turbulence effected by nonzero boresight pointing errors can be approximated at low SNR in closed-form in terms of simple elementary functions by
\begin{equation}\label{Eq:EC_Single_Low_LN_BS}
\begin{aligned}
\overline{C}\underset{\mu_{r}\,<<1}{\approxeq}&\frac{\xi^{2(1-r)}}{\left(\xi^{2}+r\right)\left(\xi^{2}+1\right)^{-r}}\,\exp\left\{\frac{r\,\sigma^{2}}{2}\left(r-1\right)\right.\\
&+\left.r\,s^{2}/\left(2\,\sigma_s^{2}\right)\,\left[1/\left(\xi^{2}+1\right)-1/\left(\xi^{2}+r\right)\right]\right\}\,c\,\mu_{r}.
\end{aligned}
\end{equation}
Similarly, for LN atmospheric turbulence under zero boresight pointing errors (i.e. for $s=0$), the asymptotic approximation to the ergodic capacity at low SNR is obtained as
\begin{equation}\label{Eq:EC_Single_Low_LN}
\overline{C}\underset{\mu_{r}\,<<1}{\approxeq}\frac{\xi^{2(1-r)}}{\left(\xi^{2}+r\right)\left(\xi^{2}+1\right)^{-r}}\,\exp\left\{\frac{r\,\sigma^{2}}{2}\left(r-1\right)\right\}\,c\,\mu_{r}.
\end{equation}
Similarly, for LN atmospheric turbulence under zero pointing errors (i.e. for $s=0$ and $\xi\rightarrow\infty$), the asymptotic approximation to the ergodic capacity at low SNR is obtained as
\begin{equation}\label{Eq:EC_Single_Low_LN_NPE}
\overline{C}\underset{\mu_{r}\,<<1}{\approxeq}\exp\left\{r\,\sigma^{2}\left(r-1\right)/2\right\}\,c\,\mu_{r}.
\end{equation}

\subsubsection{Results and Discussion}
As an illustration of the mathematical formalism presented above, simulation and numerical results for the ergodic capacity of a single FSO link transmission system under LN turbulent channels is presented as follows.

\begin{figure}[h]
\begin{center}
\includegraphics[scale=0.3]{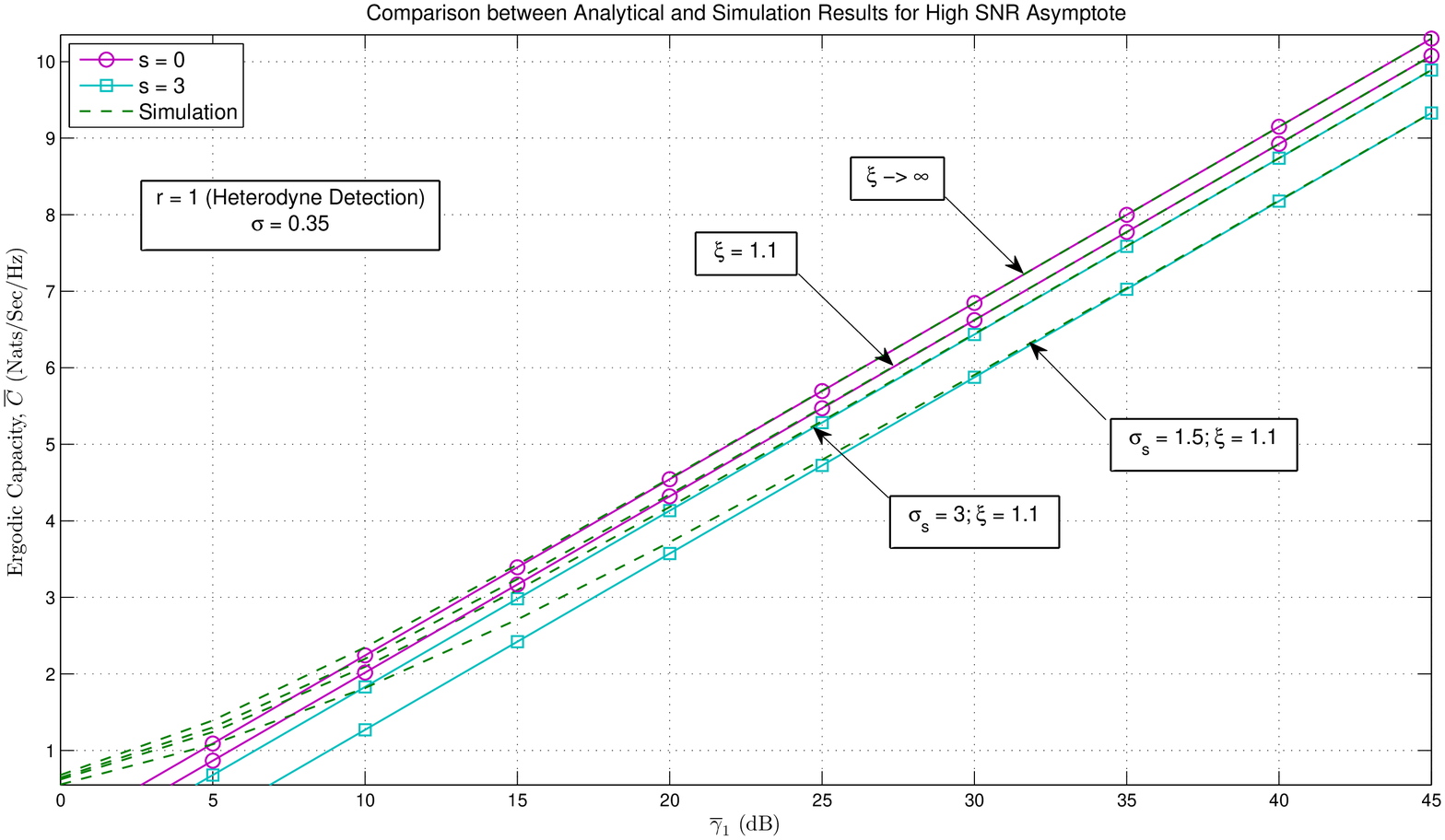}
\caption{Ergodic capacity results for varying pointing errors at high SNR regime for LN turbulence under heterodyne detection technique ($r = 1$).}
\label{Fig:EC_r2}
\end{center}
\end{figure}

\begin{figure}[h]
\begin{center}
\includegraphics[scale=0.3]{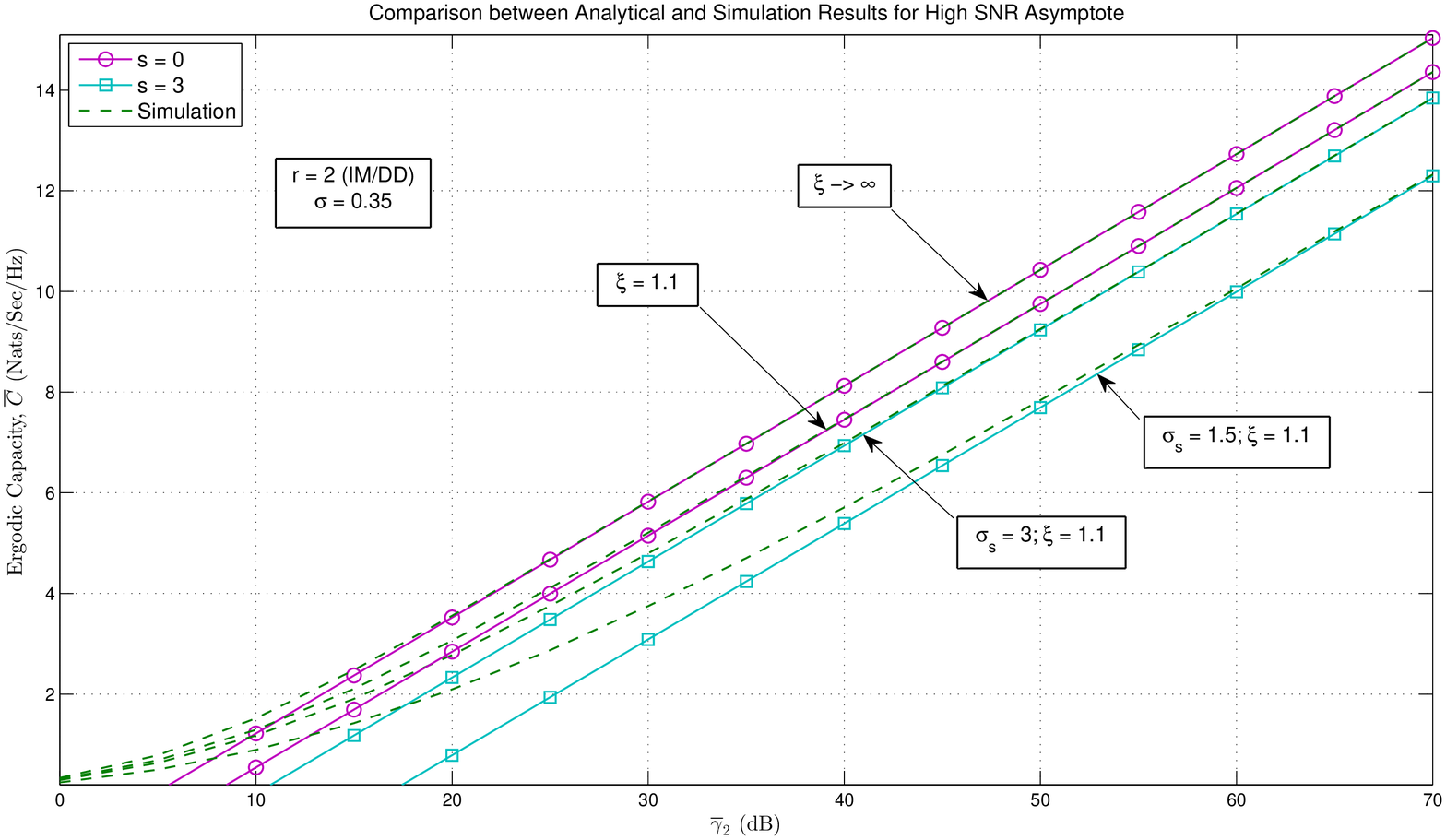}
\caption{Ergodic capacity results for varying pointing errors at high SNR regime for LN turbulence under IM/DD technique ($r = 2$).}
\label{Fig:EC_r2_2}
\end{center}
\end{figure}

\begin{figure}[h]
\begin{center}
\includegraphics[scale=0.29]{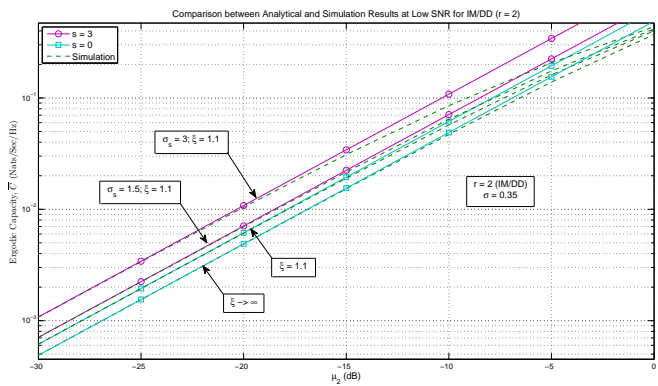}
\caption{Ergodic capacity results for varying pointing errors at low SNR regime for LN turbulence under IM/DD technique ($r = 2$).}
\label{Fig:EC_r2_Low}
\end{center}
\end{figure}

The FSO link is modeled as a LN turbulent channel with nonzero boresight pointing errors. The dotted lines marked as simulation in the figures represent the Monte-Carlo generation for the exact results to observe the asymptotic tightness of the approximated results and to prove their validity. The ergodic capacity of the FSO channel in operation under heterodyne detection technique as well as IM/DD technique is presented in \figref{Fig:EC_r2} and \figref{Fig:EC_r2_2}, respectively, for high SNR scenario.
Subsequently, the ergodic capacity of the FSO channel in operation under IM/DD technique is presented in \figref{Fig:EC_r2_Low} for low SNR scenario\,\footnote{For readers clarification, the low SNR asymptote in \eqref{Eq:EC_Single_Low_LN_BS} is actually the average SNR and hence the plot in \figref{Fig:EC_r2_Low} is against the electrical SNR.}.
These figures demonstrate the obtained results for varying effects of pointing errors with $\sigma=0.35$.\,\footnote{It is important to note here that these values for the parameters were selected from the cited references subject to the standards to prove the validity of the obtained results and hence other specific values can be used to obtain the required results by design communication engineers before deployment.}

Expectedly, for high SNR regime (i.e. \figref{Fig:EC_r2} and \figref{Fig:EC_r2_2}), as the pointing error gets severe, the ergodic capacity starts decreasing (i.e. the lower the value of $s$ and/or the higher the value of $\xi$, the higher will be the ergodic capacity). On the other hand, for low SNR regime (i.e. \figref{Fig:EC_r2_Low}), as the pointing error gets severe, the ergodic capacity starts increasing (i.e. the lower the value of $s$ and/or the higher the value of $\xi$, the lower will be the ergodic capacity).

\begin{figure}[h]
\begin{center}
\includegraphics[scale=0.3]{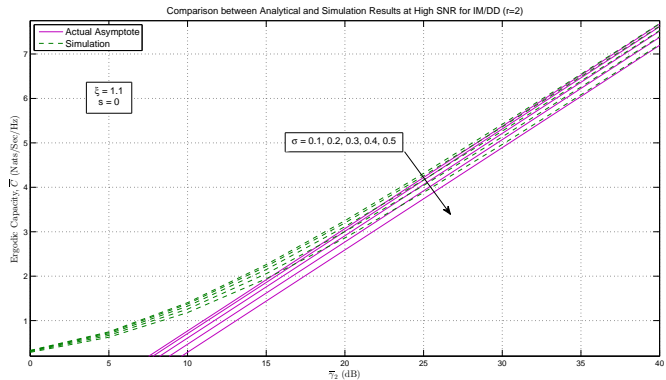}
\caption{Ergodic capacity results for IM/DD technique and varying $\sigma$ at high SNR regime for LN turbulence.}
\label{Fig:EC_LN_sigma}
\end{center}
\end{figure}

Furthermore, it can be seen that at high SNR, the asymptotic expression derived in \eqref{Eq:Moments_Derivative_n0_LN_BS} via utilizing moments gives very tight asymptotic results in high SNR regime and the same can be observed for the low SNR regime too corresponding to \eqref{Eq:EC_Single_Low_LN_BS}.
\figref{Fig:EC_LN_sigma} presents the effect of varying scintillation index parameter $\sigma=0.1, 0.2, 0.3, 0.4, 0.5$.
The pointing error effect is fixed at $s=0$ and $\xi=1.1$, and the ergodic capacity is plotted for the IM/DD technique (i.e. $r=2$). It can be observed that as the scintillation index increases, the ergodic capacity degrades.

\subsection{Rician-Lognormal (RLN) Turbulence Scenario}
\subsubsection{Exact Analysis}
For RLN atmospheric turbulence scenario under nonzero boresight pointing errors and zero boresight pointing errors, we respectively substitute \eqref{Eq:RLN_PDF_BS} and \eqref{Eq:RLN_PDF} in \eqref{Eq:EC}. To the best of our knowledge, both the above scenarios can not be solved in exact closed-form.

Additionally, we have already reached to a conclusion that it is not possible to solve the inner integral for nonzero boresight pointing errors in \eqref{Eq:EC_Int} with \eqref{Eq:PE_BS}. Hence, we end up with a three-integral expression involving the $I_{a_{R}}$ and $I_{a_{L}}$ independently. We have already learned from the previous subsection that the middle integral for LN PDF $f_{L}\left(I_{a_{L}}\right)$ in \eqref{Eq:EC_Int} with \eqref{Eq:L-N_PDF} does not lead to possible exact closed-form results and similarly the outer integral for the Rician PDF $f_{R}\left(I_{a_{R}}\right)$ in \eqref{Eq:EC_Int} with \eqref{Eq:R_PDF} also does not lead to possible exact closed-form results.

On the other hand, although we have been able to solve the inner integral for zero boresight pointing errors in \eqref{Eq:EC_Int} with \eqref{Eq:PE} to obtain \eqref{Eq:EC_Int2} but on placing the LN PDF $f_{L}\left(I_{a_{L}}\right)$ \eqref{Eq:L-N_PDF} and the Rician PDF $f_{R}\left(I_{a_{R}}\right)$ \eqref{Eq:R_PDF} into \eqref{Eq:EC_Int2}, we end up having a double integral. To the best of our knowledge, this double integral does not has an exact closed-form solution nor this double integral can be reduced further to a single integral for other possible solutions. Therefore, the ergodic capacity is analyzed utilizing the moments derived in previous section.

\subsubsection{Approximate Analysis}
Based on \eqref{Eq:EC_Single_High_Def_Simplify}, the first derivative of the moments derived in \eqref{Eq:Moments_RLN_BS} is obtained as
\begin{equation}\label{Eq:Moments_Derivative_RLN_BS}\small
\begin{aligned}
&\frac{\partial}{\partial n}\Expected{\gamma_{r}^n}=\frac{\xi^{2(1-r\,n)}\,\Gamma\left(r\,n+1\right)}{\left(\xi^{2}+r\,n\right)\left(\xi^{2}+1\right)^{-r\,n}\,\left(1+k^{2}\right)^{r\,n}}\\
&\times\exp\left\{\frac{r\,n\,\sigma^{2}}{2}\left(r\,n-1\right)+\frac{r\,n\,s^{2}}{2\,\sigma_s^{2}}\left(\frac{1}{\xi^{2}+1}-\frac{1}{\xi^{2}+r\,n}\right)\right\}\\
&\times\left\{\Hypergeom{1}{1}{-r\,n}{1}{-k^{2}}\left[-r/\left(\xi^{2}+r\,n\right)+r\,\sigma^{2}\left(\,r\,n-1/2\right)\right.\right.\\
&+r\,s^{2}/\left(2\,\sigma_s^{2}\right)\left[r\,n/\left(\xi^{2}+r\,n\right)^{2}+1/\left(\xi^{2}+1\right)-1/\left(\xi^{2}+r\,n\right)\right]\\
&-r\,\ln\left\{\xi^{2}/\left(\xi^{2}+1\right)\right\}+r\,\psi\left(r\,n+1\right)-r\,\ln\left\{k^{2}+1\right\}\\
&\left.+\ln\left\{c\,\mu_{r}\right\}\right]-\left.r\,\partial/\partial n\,\Hypergeom{1}{1}{-r\,n}{1}{-k^{2}}\right\}\left(c\,\mu_{r}\right)^{n},
\end{aligned}
\end{equation}\normalsize
where $\psi\left(.\right)$ is the digamma (psi) function \cite[Eq. (6.3.1)]{abramowitz}. It can be seen from \eqref{Eq:Moments_Derivative_RLN_BS} that the last term is in a form of derivative definition. To the best of the authors' knowledge, the derivative of $\partial/\partial a\,\Hypergeom{1}{1}{a}{b}{z}$ or $\partial/\partial b\,\Hypergeom{1}{1}{a}{b}{z}$ is not available in the open mathematical literature though this can be solved for the special case when the variable being derived with respect to, is set to $0$ i.e. $\partial/\partial a\,\Hypergeom{1}{1}{a}{b}{z}|_{a=0}$ or $\partial/\partial b\,\Hypergeom{1}{1}{a}{b}{z}|_{b=0}$ \cite[App. A]{stanford}. Hence, $\partial/\partial n\,\Hypergeom{1}{1}{-r\,n}{1}{-k^{2}}|_{n=0}$ can be solved as \cite[Eq. (38a)]{ancarani}
\begin{equation}\label{Eq:F}
\partial/\partial n\,\Hypergeom{1}{1}{-r\,n}{1}{-k^{2}}|_{n=0}=-k^{2}\,\Hypergeom{2}{2}{1,1}{2,2}{-k^{2}}.
\end{equation}
Now, substituting \eqref{Eq:F} into \eqref{Eq:Moments_Derivative_RLN_BS} and evaluating \eqref{Eq:Moments_Derivative_RLN_BS} at $n=0$, following is obtained
\begin{equation}\label{Eq:Moments_Derivative_n0_RLN_BS}\small
\begin{aligned}
&\overline{C}\underset{\mu_{r}\,>>1}{\approxeq}\ln\left\{c\,\mu_{r}\right\}-r\left[\frac{1}{\xi^{2}}+\frac{\sigma^2}{2}+\frac{s^{2}}{2\,\sigma_s^{2}\,\xi^{2}\,\left(\xi^{2}+1\right)}\right.\\
&+\left.\ln\left\{\frac{\xi^2}{\xi^2+1}\right\}+\ln\left\{k^{2}+1\right\}+\gamma_{E}-k^{2}\,\Hypergeom{2}{2}{1,1}{2,2}{-k^{2}}\right],
\end{aligned}
\end{equation}\normalsize
where $\gamma_{E}\approxeq 0.577216$ denotes the Euler-Mascheroni constant/Euler's Gamma/Euler's constant \cite{lin}. On further utilizing \cite{lin}, eq. \eqref{Eq:Moments_Derivative_n0_RLN_BS} can be simplified to
\begin{equation}\label{Eq:Moments_Derivative_n0_Alter0_RLN_BS}
\begin{aligned}
\overline{C}&\underset{\mu_{r}\,>>1}{\approxeq}\ln\left\{c\,\mu_{r}\right\}-r\left[\frac{1}{\xi^{2}}+\frac{\sigma^2}{2}+\frac{s^{2}}{2\,\sigma_s^{2}\,\xi^{2}\,\left(\xi^{2}+1\right)}\right.\\
&+\left.\ln\left\{\xi^2/\left(\xi^2+1\right)\right\}-\ln\left\{k^{2}/\left(k^{2}+1\right)\right\}-\Gamma\left(0,k^{2}\right)\right].
\end{aligned}
\end{equation}
Equation \eqref{Eq:Moments_Derivative_n0_Alter0_RLN_BS} can be further simplified via utilizing \cite[Eq. (6.5.15)]{abramowitz} to obtain
\begin{equation}\label{Eq:Moments_Derivative_n0_Alter_RLN_BS}
\begin{aligned}
\overline{C}&\underset{\mu_{r}\,>>1}{\approxeq}\ln\left\{c\,\mu_{r}\right\}-r\left[\frac{1}{\xi^{2}}+\frac{\sigma^2}{2}+\frac{s^{2}}{2\,\sigma_s^{2}\,\xi^{2}\,\left(\xi^{2}+1\right)}\right.\\
&+\left.\ln\left\{\xi^2/\left(\xi^2+1\right)\right\}-\ln\left\{k^{2}/\left(k^{2}+1\right)\right\}-E_{1}\left(k^{2}\right)\right],
\end{aligned}
\end{equation}
where $E_{n}\left(z\right)$ is an exponential integral \cite[Sec. 5.1]{abramowitz}. Hence, eq. \eqref{Eq:Moments_Derivative_n0_Alter_RLN_BS} gives the required expression for the ergodic capacity $\overline{C}$ at high SNR in terms of simple elementary functions for RLN FSO turbulent channels under the effect of boresight pointing errors. Similarly, for RLN atmospheric turbulence under zero boresight pointing errors (i.e. for $s=0$) and under no pointing errors (i.e. for $s=0$ and $\xi\rightarrow\infty$), the asymptotic approximations to their respective ergodic capacity's at high SNR are derived in \tableref{Tab:Special_Cases}.

Furthermore, for \textit{\textbf{low SNR}} asymptotic analysis, it can be easily shown that the ergodic capacity can be asymptotically approximated by the first moment. Utilizing \eqref{Eq:Moments_RLN_BS} via placing $n=1$ in it, the ergodic capacity of a single FSO link under RLN FSO turbulence effected by nonzero boresight pointing errors can be approximated at low SNR in closed-form in terms of simple elementary functions by
\begin{equation}\label{Eq:EC_RLN_BS}
\begin{aligned}
\overline{C}&\underset{\mu_{r}\,<<1}{\approxeq}\xi^{2(1-r)}/\left[\left(\xi^{2}+r\right)\left(\xi^{2}+1\right)^{-r}\right]\\
&\times\exp\left\{\frac{r\,\sigma^{2}}{2}\left(r-1\right)+\frac{r\,s^{2}}{2\,\sigma_s^{2}}\left(\frac{1}{\xi^{2}+1}-\frac{1}{\xi^{2}+r}\right)\right\}\\
&\times\Gamma\left(r+1\right)\,\Hypergeom{1}{1}{-r}{1}{-k^{2}}/\left(k^{2}+1\right)^{r}\,c\,\mu_{r}.
\end{aligned}
\end{equation}
Similarly, for RLN atmospheric turbulence under zero boresight pointing errors (i.e. for $s=0$), the asymptotic approximation to the ergodic capacity at low SNR is obtained as
\begin{equation}\label{Eq:EC_RLN}
\begin{aligned}
\overline{C}&\underset{\mu_{r}\,<<1}{\approxeq}\frac{\xi^{2(1-r)}}{\left(\xi^{2}+r\right)\left(\xi^{2}+1\right)^{-r}}\exp\left\{\frac{r\,\sigma^{2}}{2}\left(r-1\right)\right\}\\
&\times\Gamma\left(r+1\right)\,\Hypergeom{1}{1}{-r}{1}{-k^{2}}/\left(1+k^{2}\right)^{r}\,c\,\mu_{r}.
\end{aligned}
\end{equation}
Similarly, for RLN atmospheric turbulence under zero pointing errors (i.e. for $s=0$ and $\xi\rightarrow\infty$), the asymptotic approximation to the ergodic capacity at low SNR is obtained as
\begin{equation}\label{Eq:EC_RLN_NPE}\small
\overline{C}\underset{\mu_{r}\,<<1}{\approxeq}\exp\left\{\frac{r\,\sigma^{2}}{2}\left(r-1\right)\right\}\frac{\Gamma\left(r+1\right)\,\Hypergeom{1}{1}{-r}{1}{-k^{2}}}{\left(1+k^{2}\right)^{r}}\,c\,\mu_{r}.
\end{equation}\normalsize

\subsubsection{Results and Discussion}
As an illustration of the mathematical formalism presented above, simulation and numerical results for the ergodic capacity of a single FSO link transmission system under RLN turbulent channels is presented as follows.

\begin{figure}[h]
\begin{center}
\includegraphics[scale=0.3]{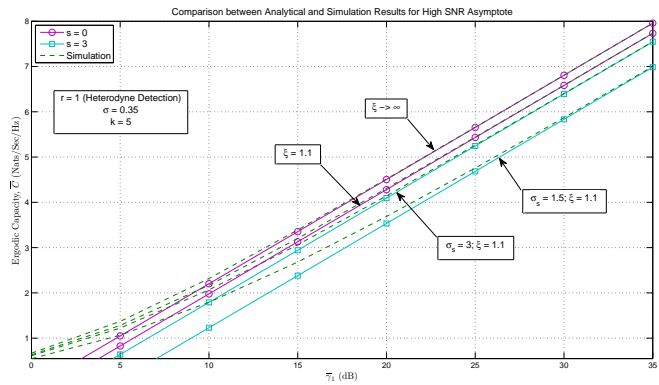}
\caption{Ergodic capacity results for varying pointing errors at high SNR regime for RLN turbulence under heterodyne detection technique ($r = 1$).}
\label{Fig:EC_RLN}
\end{center}
\end{figure}

\begin{figure}[h]
\begin{center}
\includegraphics[scale=0.3]{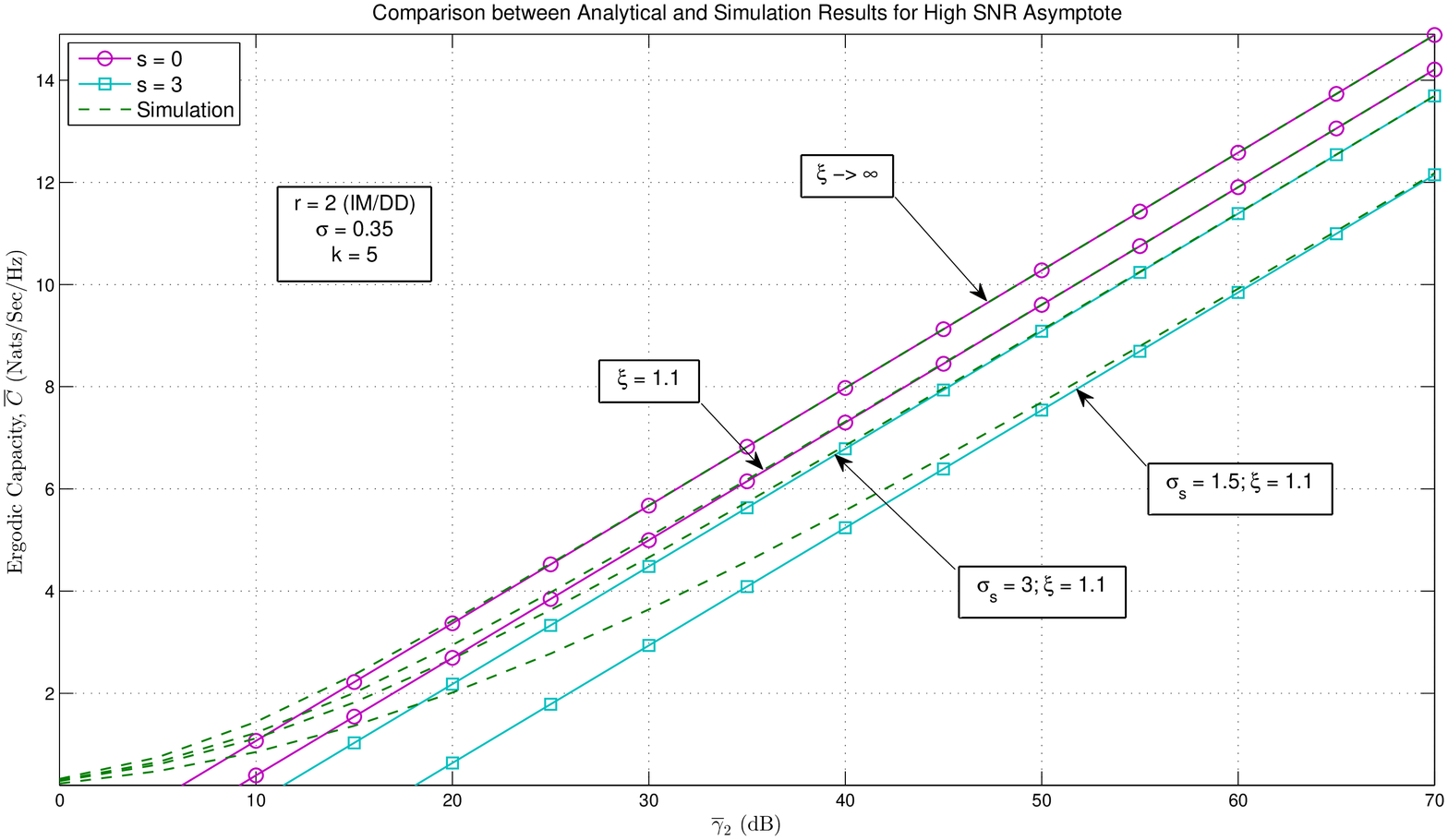}
\caption{Ergodic capacity results for varying pointing errors at high SNR regime for RLN turbulence under IM/DD technique ($r = 2$).}
\label{Fig:EC_RLN_2}
\end{center}
\end{figure}

\begin{figure}[h]
\begin{center}
\includegraphics[scale=0.29]{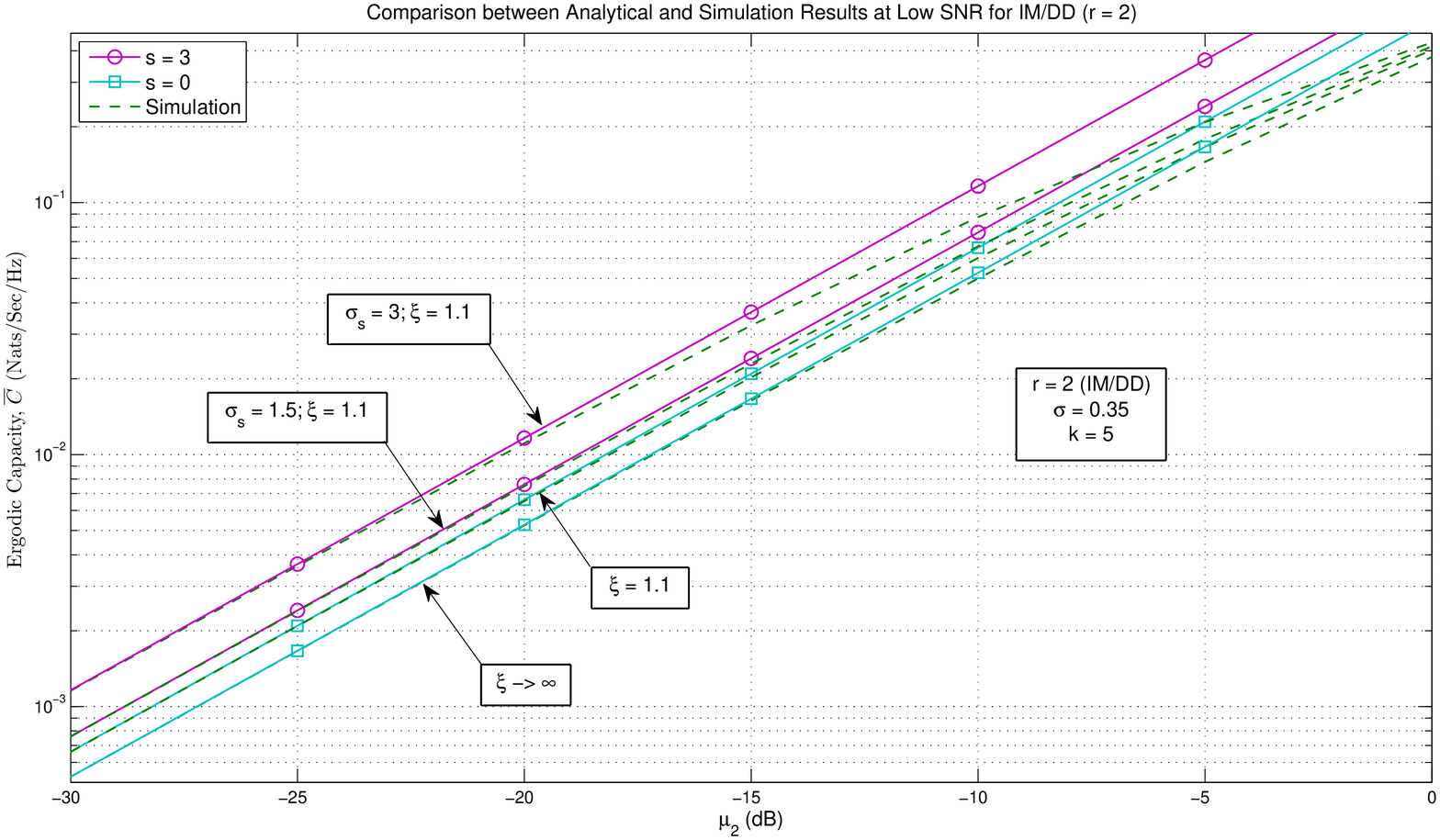}
\caption{Ergodic capacity results for varying pointing errors at low SNR regime for RLN turbulence under IM/DD technique ($r = 2$).}
\label{Fig:EC_RLN_Low}
\end{center}
\end{figure}

\begin{figure}[h]
\begin{center}
\includegraphics[scale=0.3]{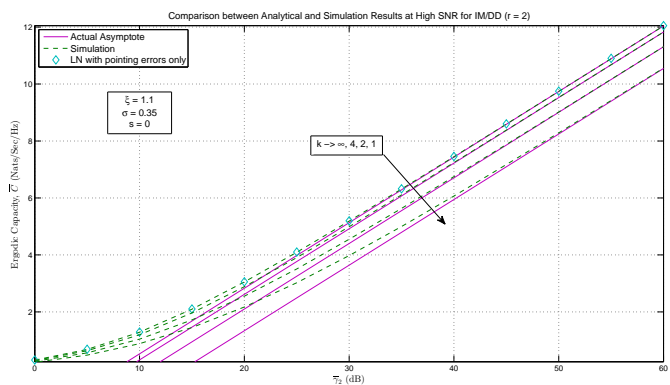}
\caption{Ergodic capacity results for IM/DD technique and varying $k$ at high SNR regime for RLN turbulence.}
\label{Fig:EC_RLN_k}
\end{center}
\end{figure}

The FSO link is modeled as composite RLN turbulent channel. The ergodic capacity of the FSO channel in operation under heterodyne detection technique as well as IM/DD technique is presented in \figref{Fig:EC_RLN} and \figref{Fig:EC_RLN_2}, respectively, for high SNR scenario.
Subsequently, the ergodic capacity of the FSO channel in operation under IM/DD technique is presented in \figref{Fig:EC_RLN_Low} for low SNR scenario\,\footnote{For readers clarification, the low SNR asymptote in \eqref{Eq:EC_RLN_BS} is actually the average SNR and hence the plot in \figref{Fig:EC_RLN_Low} is against the electrical SNR.}.
These figures demonstrate the obtained results for varying effects of pointing error with $k=5$ and $\sigma=0.35$.\,\footnote{It is important to note here that these values for the parameters were selected from the cited references subject to the standards to prove the validity of the obtained results and hence other specific values can be used to obtain the required results by design communication engineers before deployment.} Similar trend in results can be observed here as were observed for the LN only scenario in \figref{Fig:EC_r2}, \figref{Fig:EC_r2_2}, and \figref{Fig:EC_r2_Low}. \figref{Fig:EC_RLN_k} presents the effect of varying $k$ turbulence parameter $k \rightarrow \infty, 4, 2, 1$.
The pointing error effect is fixed at $s=0$ and $\xi=1.1$, and the LN scintillation index is fixed at $\sigma=0.35$. The ergodic capacity is plotted for the IM/DD technique (i.e. $r=2$). It can be observed that as the turbulence parameter $k$ increases, the ergodic capacity improves and ultimately matches with LN turbulence (signified with a diamond shape symbol in \figref{Fig:EC_RLN_k}) as $k\rightarrow\infty$ (i.e. Rician turbulence becomes negligible).

Moreover, it is important to note that these plots are very useful to easily obtain the approximation error of the asymptotic results obtained by the proposed moments-based approximation method or in other words to find the accuracy of the proposed moments-based approximation method. For instance, let us refer to the third curve from the top that corresponds to $s=3$, $\sigma_{s}=3$, and $\xi=1.1$ in \figref{Fig:EC_RLN_2}. Let us assume that we want to control the approximation error to, lets say, around $3.9\%$ or less. Now, we can easily deduce the channel performance i.e. at $\overline{\gamma}_2=30$ dB; $\overline{C}=\{4.66~(\rm{exact})$, $4.482~(\rm{simulation})\}$ with approximation error $=3.8197\%$. Based on this, we can easily conclude that for an acceptable approximation error of $3.9 \%$ or less, our average SNR must be at least $\overline{\gamma}_2=30$ dB or more. Similarly, if we want to look at this in another way i.e. our system is operating at a certain average SNR and we would like find out the accuracy of our approximation then this can also be obtained easily as follows. We can easily deduce that at $\overline{\gamma}_2=30$ dB, $\overline{C}=\{4.66~(\rm{exact})$, $4.482~(\rm{simulation})\}$ that leads to an approximation error $=3.8197\%$. Similarly, at $\overline{\gamma}_2=35$ dB; $\overline{C}=\{5.741~(\rm{exact})$, $5.633~(\rm{simulation})\}$ leads to an approximation error $=1.8812\%$, and at $\overline{\gamma}_2=40$ dB; $\overline{C}=\{6.849~(\rm{exact})$, $6.784~(\rm{simulation})\}$ leads to an approximation error $=0.949\%$.

\subsection{M\'{a}laga ($\mathcal{M}$) Turbulence Scenario}
\subsubsection{Exact Analysis}
For $\mathcal{M}$ atmospheric turbulence scenario under nonzero boresight pointing errors, we respectively substitute \eqref{Eq:M_PDF_BS} and \eqref{Eq:M_PDF} in \eqref{Eq:EC}. To the best of our knowledge, both the above scenarios can not be solved in exact closed-form.

Additionally, we have already reached to a conclusion that it is not possible to solve the inner integral for nonzero boresight pointing errors in \eqref{Eq:EC_Int} with \eqref{Eq:PE_BS}. Hence, we end up with a double-integral expression involving the $I_{a_{M}}$. The integral with respect to $I_{a_{M}}$ can be solved in exact closed-form to obtain
\begin{equation}\label{Eq:EC_Int2_M_BS}
\begin{aligned}
\overline{C}&=\int_{0}^{A_{0}}\int_{0}^{\infty}\,\ln\left\{1+\frac{c\left(\eta_{e}\,I_{l}\,I_{a_{M}}\,I_{p}\right)^{r}}{N_{0}}\right\}\,f_M\left(I_{a_{M}}\right)\,f_p\left(I_{p}\right)\\
&\times dI_{a_{M}}\,dI_{p}=\xi^{2}\,A\,r^{3}/\left[2\,A_{0}^{\xi^{2}}\,\left(2\,\pi\right)^{r-1}\right]\\
&\times\left[\left(g\,\beta+\Omega^{'}\right)/\left(\alpha\,\beta\right)\right]^{2}\,\exp\left\{-s^{2}/\left(2\,\sigma_s^{2}\right)\right\}\\
&\times\sum_{m=1}^{\beta}a_{m}\int_{0}^{A_{0}}I_{p}^{\xi^{2}-1}\,\BesselI{s/\sigma_s\,\sqrt{-2/\xi^{-2}\,\ln\left\{I_{p}/A_{0}\right\}}}\\
&\times\MeijerG[right]{1,2\,r+2}{2\,r+2,2}{\frac{c\left(\eta_{e}\,I_{l}\,I_{p}\right)^{r}\,\left(g\,\beta+\Omega^{'}\right)^{r}}{r^{-2\,r}\,N_{0}\,\left(\alpha\,\beta\right)^{r}}}{1,1,\kappa_{0}}{1,0}\,dI_{p},
\end{aligned}
\end{equation}
where $\kappa_{0}=\frac{-1-\left(\alpha-m\right)/2}{r},\dots,\frac{-2-\left(\alpha-m\right)/2+r}{r},\frac{-1-\left(m-\alpha\right)/2}{r},$ $\dots,\frac{-2-\left(m-\alpha\right)/2+r}{r}$ comprises of $2r$ terms. To the best of our knowledge, this single integral in \eqref{Eq:EC_Int2_M_BS} does not have an exact closed-form solution\,\footnote{Please note that similar integral results/outcomes were obtained for GG turbulence scenario under nonzero boresight pointing errors.}.

On the other hand, for $\mathcal{M}$ atmospheric turbulence under zero boresight pointing errors, utilizing \eqref{Eq:EC} by placing \eqref{Eq:M_PDF} in it results into an exact closed-form result as \cite[Eq. (20)]{ansari12}
\begin{equation}\label{Eq:EC_Single_M}
\overline{C}=\frac{D}{\ln(2)}\sum_{m=1}^{\beta}c_{m}\,\MeijerG[right]{3r+2,1}{r+2,3r+2}{\frac{E}{c\,\mu_{r}}}{0,1,\kappa_1}{\kappa_2,0,0},
\end{equation}
where $D=\xi^{2}A/\left[2^{r}(2\,\pi)^{r-1}\right]$, $c_{m}=a_{m}\,b_{m}\,r^{\alpha+m-1}$, $E=\left(B\,\xi^{2}\right)^{\,r}/\left[\left(\xi^{2}+1\right)^{r}r^{2\,r}\right]$, $\kappa_1=\frac{\xi^2+1}{r},\dots,\frac{\xi^2+r}{r}$ comprises of $r$ terms, and $\kappa_2=\frac{\xi^2}{r},\dots,\frac{\xi^2+r-1}{r},\frac{\alpha}{r},\dots,\frac{\alpha+r-1}{r}, \frac{m}{r},\dots,\frac{m+r-1}{r}$ comprises of $3r$ terms.
Similarly, as a special case, an exact closed-form result for the moments of GG atmospheric turbulence under zero boresight pointing errors is obtained as \cite[Eq. (13)]{ansari11}
\begin{equation}\label{Eq:EC_Single_G}
\overline{C}=\frac{J}{\ln(2)}\,\MeijerG[right]{3r+2,1}{r+2,3r+2}{\frac{K}{c\,\mu_{r}}}{0,1,\kappa_1}{\kappa_3,0,0},
\end{equation}
where $J=r^{\alpha+\beta-2}\,\xi^2/\left[(2\,\pi)^{r-1}\,\Gamma(\alpha)\,\Gamma(\beta)\right]$, $K=(\xi^{2}\alpha\,\beta)^r/\left[\left(\xi^{2}+1\right)^{r}r^{2\,r}\right]$, and $\kappa_3=\frac{\xi^2}{r},\dots,\frac{\xi^2+r-1}{r},\frac{\alpha}{r},\dots,\frac{\alpha+r-1}{r}, \frac{\beta}{r},\dots,\frac{\beta+r-1}{r}$ comprises of $3r$ terms.

\subsubsection{Approximate Analysis}
Reverting back to $\mathcal{M}$ atmospheric turbulence under nonzero boresight pointing errors, since it is not feasible to obtain an exact closed-form solution, we utilize the moments derived earlier to deduce the asymptotic results. Hence, based on \eqref{Eq:EC_Single_High_Def_Simplify}, the first derivative of the moments in \eqref{Eq:Moments_M_BS} is required to be evaluated at $n=0$ for high SNR asymptotic approximation to the ergodic capacity. The first derivative of the moments in \eqref{Eq:Moments_M_BS} is given as
\begin{equation}\label{Eq:Moments_Derivative_M_BS}\small
\begin{aligned}
&\frac{\partial}{\partial n}\Expected{\gamma_{r}^n}=\frac{\xi^{2\left(1-r\,n\right)}\,r\,A\,\Gamma\left(r\,n+\alpha\right)}{\left(\xi^{2}+r\,n\right)\,\left(\xi^{2}+1\right)^{-r\,n}\,2^{r}\,B^{r\,n}}\sum_{m=1}^{\beta}b_{m}\,\Gamma\left(r\,n+m\right)\\
&\times\,\exp\left\{r\,n\,s^{2}/\left(2\,\sigma_s^{2}\right)\left[1/\left(\xi^{2}+1\right)-1/\left(\xi^{2}+r\,n\right)\right]\right\}\\
&\times\left[-r/\left(\xi^{2}+r\,n\right)-r\,\ln\left\{\xi^{2}/\left(\xi^{2}+1\right)\right\}-r\,\ln\left\{B\right\}\right.\\
&+r\,s^{2}/\left(2\,\sigma_s^{2}\right)\left[r\,n/\left(\xi^{2}+r\,n\right)^{2}+1/\left(\xi^{2}+1\right)-1/\left(\xi^{2}+r\,n\right)\right]\\
&\left.+r\,\psi\left(r\,n+\alpha\right)+r\,\psi\left(r\,n+m\right)+\ln\left\{c\,\mu_{r}\right\}\right]\,\left(c\,\mu_{r}\right)^{n},\\
\end{aligned}
\end{equation}\normalsize
and at $n=0$, it evaluates to
\begin{equation}\label{Eq:Moments_Derivative_n0_M_BS}
\begin{aligned}
&\overline{C}\underset{\mu_{r}\,>>1}{\approxeq}\frac{r\,A\,\Gamma(\alpha)}{2^{r}}\sum_{m=1}^{\beta}b_{m}\,\Gamma(m)\,\left\{r\left[-1/\xi^2-\ln(B)+\psi(\alpha)\right.\right.\\
&-\left.\left.\frac{s^{2}\,\sigma_s^{-2}}{2\,\xi^{2}\,\left(\xi^{2}+1\right)}-\ln\left\{\frac{\xi^{2}}{\xi^{2}+1}\right\}+\psi(m)\right]+\ln(c\,\mu_{r})\right\}.
\end{aligned}
\end{equation}
For GG atmospheric turbulence, as a special case to $\mathcal{M}$ turbulence, the first derivative, evaluated at $n=0$, of the moments in \eqref{Eq:Moments_G_BS} is derived in \tableref{Tab:Special_Cases}.
Now, for $\mathcal{M}$ and GG atmospheric turbulences under zero boresight pointing errors (i.e. for $s=0$) and under zero pointing errors (i.e. for $s=0$ and $\xi\rightarrow\infty$), the asymptotic approximations to the respective ergodic capacity's at high SNR are derived in \tableref{Tab:Special_Cases}.
Alternatively, for $\mathcal{M}$ and GG atmospheric turbulences under zero boresight pointing errors (i.e. for $s=0$), the ergodic capacity's in \eqref{Eq:EC_Single_M} and \eqref{Eq:EC_Single_G} can be expressed asymptotically via utilizing the Meijer's G function expansion as \cite[Eq. (17)]{ansari11}
\begin{equation}\label{Eq:CAP_Asymp_M}
\begin{aligned}
\overline{C}&\underset{\mu_{r}\,>>1}{\approxeq}\frac{D}{\ln(2)}\sum_{m=1}^{\beta}c_{m}\sum_{k=1}^{3r+2}\left(\frac{c\,\mu_{r}}{E}\right)^{-\kappa_{2,k}}\\
&\times\frac{\Gamma(1+\kappa_{2,k})\prod_{l=1;\,l\neq k}^{3r+2}\Gamma(\kappa_{2,l}-\kappa_{2,k})}{\Gamma(1-\kappa_{2,k})\prod_{l=1}^{r}\Gamma(\kappa_{1,l}-\kappa_{2,k})},
\end{aligned}
\end{equation}
and
\begin{equation}\label{Eq:CAP_Asymp_G}
\begin{aligned}
\overline{C}&\underset{\mu_{r}\,>>1}{\approxeq}\frac{A}{\ln(2)}\sum_{k=1}^{3r+2}\left(\frac{c\,\mu_{r}}{B}\right)^{-\kappa_{3,k}}\\
&\times\frac{\Gamma(1+\kappa_{3,k})\prod_{l=1;\,l\neq k}^{3r+2}\Gamma(\kappa_{3,l}-\kappa_{3,k})}{\Gamma(1-\kappa_{3,k})\prod_{l=1}^{r}\Gamma(\kappa_{1,l}-\kappa_{3,k})},
\end{aligned}
\end{equation}
respectively, where $\kappa_{u,v}$ represents the $v^{\mathrm{th}}$-term of $\kappa_{u}$.

Furthermore, for \textit{\textbf{low SNR}} asymptotic analysis, it can be easily shown that the ergodic capacity can be asymptotically approximated by the first moment. Utilizing \eqref{Eq:Moments_M_BS} and \eqref{Eq:Moments_G_BS} via placing $n=1$ in them, the ergodic capacity's of a single FSO link under $\mathcal{M}$ and GG FSO turbulences effected by nonzero boresight pointing errors can be approximated at low SNR in closed-form in terms of simple elementary functions by
\begin{equation}\label{Eq:EC_M_BS}
\begin{aligned}
\overline{C}&\underset{\mu_{r}\,<<1}{\approxeq}\xi^{2(1-r)}/\left[\left(\xi^{2}+r\right)\left(\xi^{2}+1\right)^{-r}\right]\\
&\times\,\exp\left\{r\,s^{2}/\left(2\,\sigma_s^{2}\right)\left[1/\left(\xi^{2}+1\right)-1/\left(\xi^{2}+r\right)\right]\right\}\\
&\times r\,A\,\Gamma(r+\alpha)/\left(2^{r}\,B^{r}\right)\sum_{m=1}^{\beta}\,b_{m}\,\Gamma(r+m)\,c\,\mu_{r},
\end{aligned}
\end{equation}
and
\begin{equation}\label{Eq:EC_G_BS}
\begin{aligned}
\overline{C}&\underset{\mu_{r}\,<<1}{\approxeq}\frac{\xi^{2\left(1-r\right)}\,\left(\xi^{2}+1\right)^{r}\,\Gamma\left(r+\alpha\right)\,\Gamma\left(r+\beta\right)}{\left(\xi^{2}+r\right)\,\left(\alpha\,\beta\right)^{r}\,\Gamma\left(\alpha\right)\,\Gamma\left(\beta\right)}\\
&\times\,\exp\left\{r\,s^{2}/\left(2\,\sigma_s^{2}\right)\left[1/\left(\xi^{2}+1\right)-1/\left(\xi^{2}+r\right)\right]\right\}\,c\,\mu_{r},
\end{aligned}
\end{equation}
respectively. Similarly, for $\mathcal{M}$ and GG atmospheric turbulences under zero boresight pointing errors (i.e. for $s=0$), the asymptotic approximations to the ergodic capacity's at low SNR are obtained, respectively, as
\begin{equation}\label{Eq:EC_M}
\begin{aligned}
\overline{C}&\underset{\mu_{r}\,<<1}{\approxeq}\xi^{2(1-r)}/\left[\left(\xi^{2}+r\right)\left(\xi^{2}+1\right)^{-r}\right]\\
&\times r\,A\,\Gamma(r+\alpha)/\left(2^{r}\,B^{r}\right)\sum_{m=1}^{\beta}\,b_{m}\,\Gamma(r+m)\,c\,\mu_{r},
\end{aligned}
\end{equation}
and
\begin{equation}\label{Eq:EC_G}
\begin{aligned}
\overline{C}&\underset{\mu_{r}\,<<1}{\approxeq}\frac{\xi^{2\left(1-r\right)}\,\left(\xi^{2}+1\right)^{r}\,\Gamma\left(r+\alpha\right)\,\Gamma\left(r+\beta\right)}{\left(\xi^{2}+r\right)\,\left(\alpha\,\beta\right)^{r}\,\Gamma\left(\alpha\right)\,\Gamma\left(\beta\right)}\,c\,\mu_{r}.
\end{aligned}
\end{equation}
Similarly, for $\mathcal{M}$ and GG atmospheric turbulences under zero pointing errors (i.e. for $s=0$ and $\xi\rightarrow\infty$), the asymptotic approximations to the ergodic capacity's at low SNR are obtained, respectively, as
\begin{equation}\label{Eq:EC_M_NPE}
\overline{C}\underset{\mu_{r}\,<<1}{\approxeq}r\,A\,\Gamma(r+\alpha)/\left(2^{r}\,B^{r}\right)\sum_{m=1}^{\beta}\,b_{m}\,\Gamma(r+m)\,c\,\mu_{r},
\end{equation}
and
\begin{equation}\label{Eq:EC_G_NPE}
\overline{C}\underset{\mu_{r}\,<<1}{\approxeq}\frac{\Gamma\left(r+\alpha\right)\,\Gamma\left(r+\beta\right)}{\left(\alpha\,\beta\right)^{r}\,\Gamma\left(\alpha\right)\,\Gamma\left(\beta\right)}\,c\,\mu_{r}.
\end{equation}

\subsubsection{Results and Discussion}
As an illustration of the mathematical formalism presented above, simulation and numerical results for the ergodic capacity of a single FSO link transmission system under $\mathcal{M}$ turbulence channels is presented as follows. The FSO link is modeled as $\mathcal{M}$ turbulence channel with the effects of atmosphere as ($\alpha=2.296$; $\beta=2$), ($\alpha=4.2$; $\beta=3$) and ($\alpha=8$; $\beta=4$), ($\Omega=1.3265$, $b_{0}=0.1079$), $\rho=0.596$, and $\phi_{A}-\phi_{B}=\pi/2$ unless stated otherwise.\,\footnote{It is important to note here that these values for the parameters were selected from \cite{navas} subject to the standards to prove the validity of the obtained results and hence other specific values can be used to obtain the required results by design communication engineers before deployment. Also, for all cases, $10^{6}$ realizations of the random variable were generated to perform the Monte-Carlo simulations in MATLAB.} In MATLAB, a $\mathcal{M}$ turbulent channel random variable was generated via squaring the absolute value of a Rician-shadowed random variable \cite{navas}.

\begin{figure}[h]
\begin{center}
\includegraphics[scale=0.3]{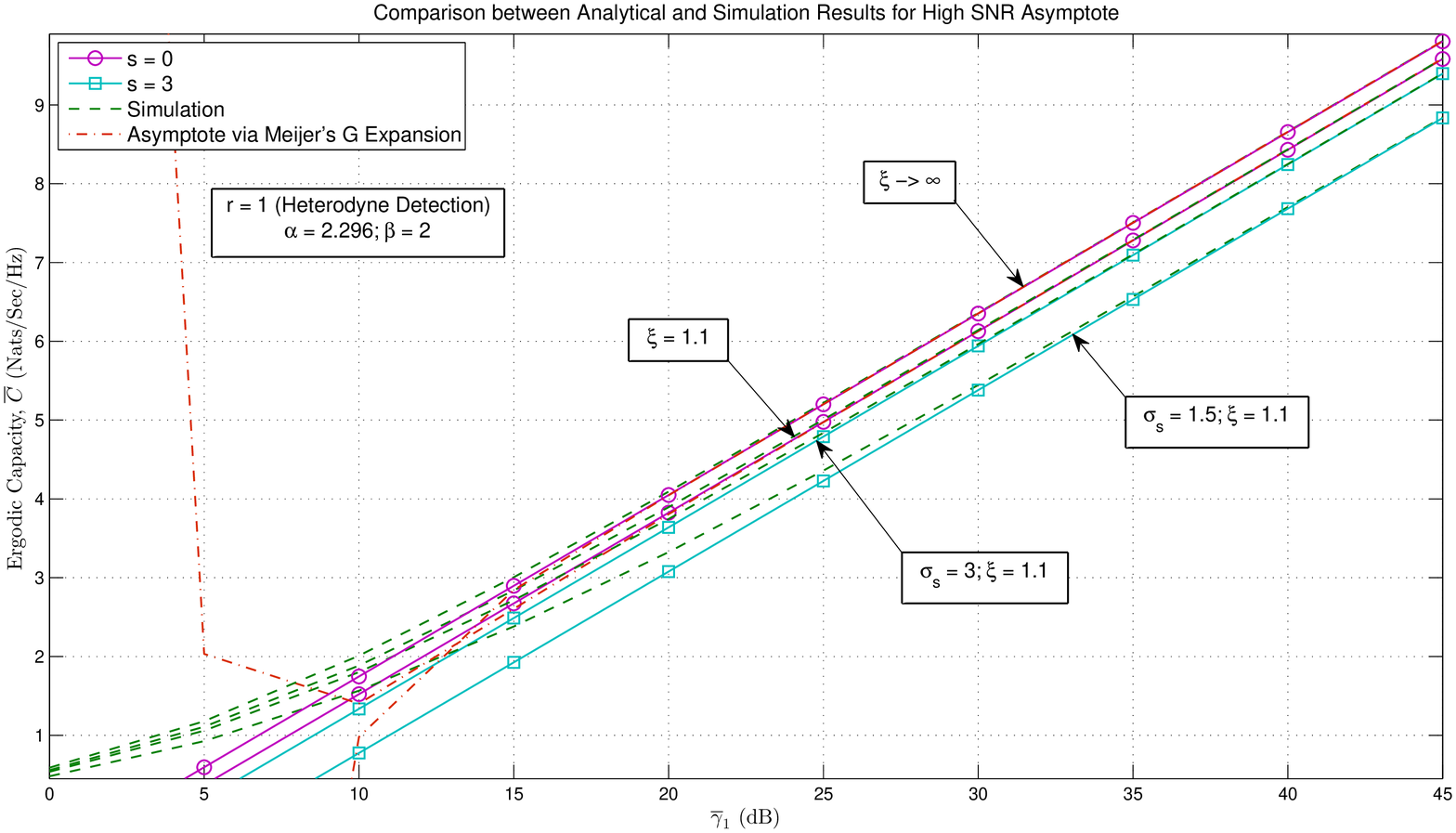}
\caption{Ergodic capacity results for varying pointing errors at high SNR regime for $\mathcal{M}$ turbulence under heterodyne detection technique ($r = 1$).}
\label{Fig:EC_M}
\end{center}
\end{figure}

\begin{figure}[h]
\begin{center}
\includegraphics[scale=0.3]{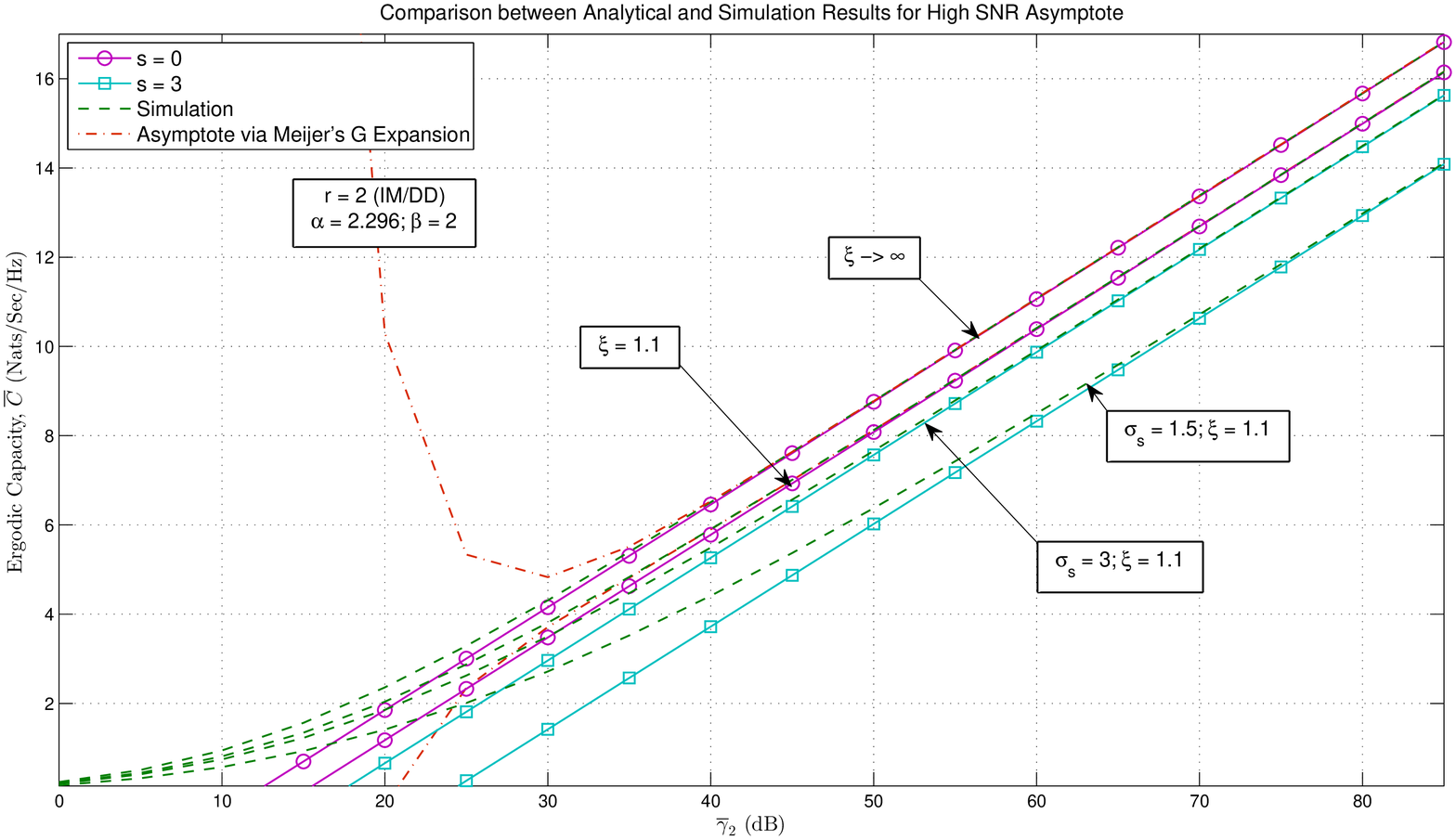}
\caption{Ergodic capacity results for varying pointing errors at high SNR regime for $\mathcal{M}$ turbulence under IM/DD technique ($r = 2$).}
\label{Fig:EC_M_2}
\end{center}
\end{figure}

\begin{figure}[h]
\begin{center}
\includegraphics[scale=0.29]{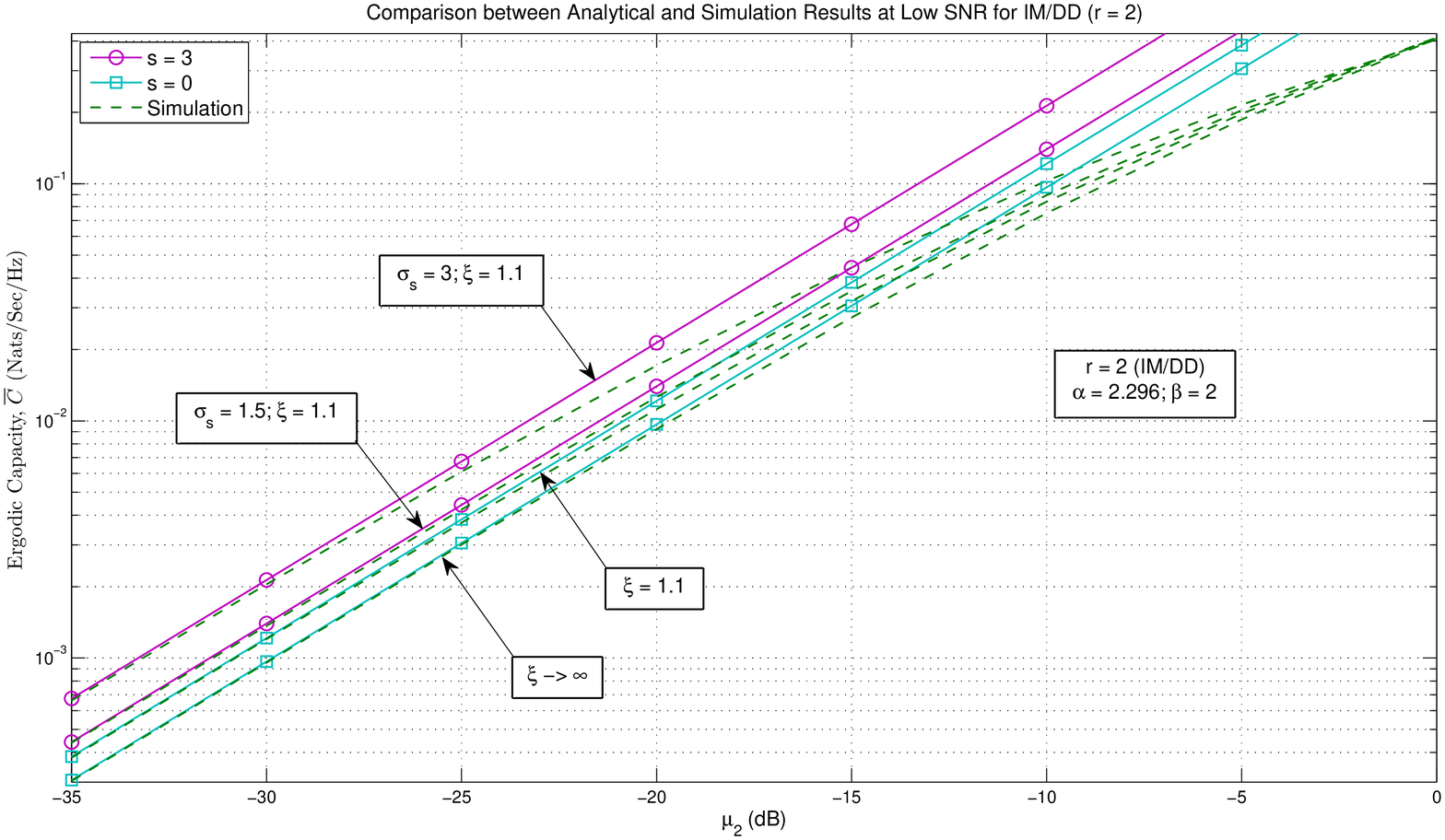}
\caption{Ergodic capacity results for varying pointing errors at low SNR regime for $\mathcal{M}$ turbulence under IM/DD technique ($r = 2$).}
\label{Fig:EC_M_Low}
\end{center}
\end{figure}

\begin{figure}[h]
\begin{center}
\includegraphics[scale=0.3]{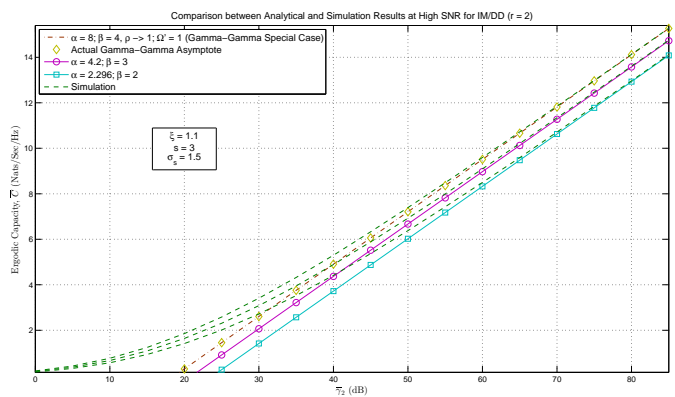}
\caption{Ergodic capacity results for IM/DD technique and varying atmospheric turbulence effects at high SNR regime for $\mathcal{M}$ turbulence.}
\label{Fig:EC_M_ab}
\end{center}
\end{figure}

The ergodic capacity of the FSO channel in operation under heterodyne detection technique as well as IM/DD technique is presented in \figref{Fig:EC_M} and \figref{Fig:EC_M_2}, respectively, for high SNR scenario.
Subsequently, the ergodic capacity of the FSO channel in operation under IM/DD technique is presented in \figref{Fig:EC_M_Low} for low SNR scenario\,\footnote{For readers clarification, the low SNR asymptote in \eqref{Eq:EC_M_BS} is actually the average SNR and hence the plot in \figref{Fig:EC_M_Low} is against the electrical SNR.}.
These figures demonstrate the obtained results for varying effects of pointing error with $\alpha=2.296$ and $\beta=2$. Similar trend in results can be observed here as were observed for the LN only and the RLN scenarios in \figref{Fig:EC_r2}, \figref{Fig:EC_r2_2}, \figref{Fig:EC_r2_Low}, \figref{Fig:EC_RLN}, \figref{Fig:EC_RLN_2}, and \figref{Fig:EC_RLN_Low}. Additionally, we have plotted in \figref{Fig:EC_M} and \figref{Fig:EC_M_2} the new Meijer's G function expansion based ergodic capacity approximate for the zero boresight pointing error case under the $\mathcal{M}$ turbulence scenario where the exact closed-form ergodic capacity involves the Meijer's G function that is given in \eqref{Eq:CAP_Asymp_M}. The plots confirm that both the approaches i.e. the moments-based approach and the Meijer's G function expansion based approach provide similar results for the ergodic capacity of such FSO atmospheric turbulence channel as the curves from both these approaches overlap simultaneously with the simulation curves nearly at a similar average SNR. \figref{Fig:EC_M_ab} presents the effect of varying atmospheric turbulences (i.e. varying $\alpha$'s and $\beta$'s).
The pointing error effect is fixed at $s=3$, $\sigma_{s}=1.5$, and $\xi=1.1$. The ergodic capacity is plotted for the IM/DD technique (i.e. $r=2$). It can be observed that as the turbulence gets severs, the ergodic capacity degrades and vice versa. An important observation is that we can observe that once we apply $\rho\rightarrow 1$ and $\Omega^{'}=1$, the $\mathcal{M}$ turbulence matches exactly the special case of Gamma-Gamma turbulence. This can be depicted from the case wherein ($\alpha=8$; $\beta=4$).

\subsection{Important Outcomes and Further Motivations}
\begin{itemize}
\item Hence, eqs. \eqref{Eq:Moments_Derivative_n0_LN_BS}, \eqref{Eq:Moments_Derivative_n0_Alter_RLN_BS}, and \eqref{Eq:Moments_Derivative_n0_M_BS} give the required expressions for the ergodic capacity $\overline{C}$ at high SNR in terms of simple elementary functions.
\item Some special cases of these ergodic capacity results are presented in \tableref{Tab:Special_Cases}.
\item Furthermore, at high SNR, the ergodic capacity for the optimal rate adaptation (ORA) policy and the optimal joint power and rate adaptation (OPRA) policy perform similarly. Therefore, these ergodic capacity results are applicable to both the ergodic capacity policies (i.e. ORA as well as OPRA).
\item Interestingly, the low SNR asymptotic ergodic capacity for the heterodyne detection technique (i.e. $r=1$ case) in \eqref{Eq:EC_Single_Low_LN_BS}-\eqref{Eq:EC_Single_Low_LN_NPE}, \eqref{Eq:EC_RLN_BS}-\eqref{Eq:EC_RLN_NPE}, and \eqref{Eq:EC_M_BS}-\eqref{Eq:EC_G_NPE} is actually the average SNR i.e. $\overline{C}\underset{\mu_{1}\,<<1}{\approxeq}\overline{\gamma}_{1}=\mu_{1}$.
\end{itemize}

\section{Concluding Remarks}
Unified expression for the moments of the average SNR of a FSO link operating over the LN, the RLN, and the $\mathcal{M}$ atmospheric turbulences under nonzero and zero boresight pointing errors were derived. Capitalizing on these expressions, we presented new unified asymptotic formulas applicable in high and low SNR regimes for the ergodic capacity in terms of simple elementary functions for the respective turbulence models. Subsequently, some special cases were also summarized in \tableref{Tab:Special_Cases}. In addition, this work presented simulation examples to validate and illustrate the mathematical formulations developed in this work and to show the effect of the scintillation index, the pointing errors, and the respective turbulence parameters severity on the system performance.

\bibliographystyle{IEEEtran}
\bibliography{References}

\begin{landscape}
\begin{table}
\centering\caption{Special Cases for LN, RLN, and $\mathcal{M}$ Atmospheric Turbulent High SNR Ergodic Capacities}
\label{Tab:Special_Cases}
\begin{tabular}{l c c c}\hline\hline\\[-2mm]
\textbf{Turbulence Model} & \textbf{With Nonzero Boresight Pointing Errors} & \textbf{With Zero Boresight Pointing Errors $\left(s=0\right)$} & \textbf{Without Pointing Errors $\left(s=0; \xi\rightarrow\infty\right)$}\\[0.5mm]\hline\\[-2.5mm]
Lognormal (LN) & $\ln\left\{c\,\mu_{r}\right\}-r\left[\frac{1}{\xi^{2}}+\frac{\sigma^2}{2}+\frac{s^{2}}{2\,\sigma_s^{2}\,\xi^{2}\,\left(\xi^{2}+1\right)}+\ln\left\{\frac{\xi^2}{\xi^2+1}\right\}\right]$ & $\ln\left\{c\,\mu_{r}\right\}-r\left[\frac{1}{\xi^{2}}+\frac{\sigma^2}{2}+\ln\left\{\frac{\xi^2}{\xi^2+1}\right\}\right]$ & $\ln\left\{c\,\mu_{r}\right\}-r\,\frac{\sigma^2}{2}$\\[1mm]\\[-2.5mm]
$\left(k\rightarrow\infty\right)$ &  & \\[1mm]\hline\\[-2.5mm]
Rician-LN (RLN) & $\ln\left\{c\,\mu_{r}\right\}-r\left[\frac{1}{\xi^{2}}+\frac{\sigma^2}{2}+\frac{s^{2}}{2\,\sigma_s^{2}\,\xi^{2}\,\left(\xi^{2}+1\right)}\right.$ & $\ln\left\{c\,\mu_{r}\right\}-r\left[\frac{1}{\xi^{2}}+\frac{\sigma^2}{2}+\ln\left\{\frac{\xi^2}{\xi^2+1}\right\}\right.$ & $\ln\left\{c\,\mu_{r}\right\}-r\left[\frac{\sigma^2}{2}-\ln\left\{\frac{k^{2}}{1+k^{2}}\right\}-E_{1}\left(k^{2}\right)\right]$\\[1mm]\\[-2.5mm]
 & $\left.+\ln\left\{\frac{\xi^2}{\xi^2+1}\right\}-\ln\left\{\frac{k^{2}}{k^{2}+1}\right\}-E_{1}\left(k^{2}\right)\right]$ & $\left.-\ln\left\{\frac{k^{2}}{1+k^{2}}\right\}-E_{1}\left(k^{2}\right)\right]$ & \\[1.2mm]\hline\\[-2.5mm]
Rician & $\ln\left\{c\,\mu_{r}\right\}-r\left[\frac{1}{\xi^{2}}+\frac{s^{2}}{2\,\sigma_s^{2}\,\xi^{2}\,\left(\xi^{2}+1\right)}\right.$ & $\ln\left\{c\,\mu_{r}\right\}-r\left[\frac{1}{\xi^{2}}+\ln\left\{\frac{\xi^2}{\xi^2+1}\right\}\right.$ & $\ln\left\{c\,\mu_{r}\right\}-r\left[\ln\left\{\frac{1+k^{2}}{k^{2}}\right\}-E_{1}\left(k^{2}\right)\right]$\\[1mm]\\[-2.5mm]
$\left(\sigma\rightarrow 0\right)$ & $\left.+\ln\left\{\frac{\xi^2}{\xi^2+1}\right\}-\ln\left\{\frac{k^{2}}{k^{2}+1}\right\}-E_{1}\left(k^{2}\right)\right]$ & $\left.-\ln\left\{\frac{k^{2}}{1+k^{2}}\right\}-E_{1}\left(k^{2}\right)\right]$ & \\[1.2mm]\hline\\[-2.5mm]
Rayleigh-LN & $\ln\left\{c\,\mu_{r}\right\}-r\left[\frac{1}{\xi^{2}}+\frac{\sigma^2}{2}+\frac{s^{2}}{2\,\sigma_s^{2}\,\xi^{2}\,\left(\xi^{2}+1\right)}+\ln\left\{\frac{\xi^2}{\xi^2+1}\right\}+\gamma_{E}\right]$ & $\ln\left\{c\,\mu_{r}\right\}-r\left[\frac{1}{\xi^{2}}+\frac{\sigma^2}{2}+\ln\left\{\frac{\xi^2}{\xi^2+1}\right\}+\gamma_{E}\right]$ & $\ln\left\{c\,\mu_{r}\right\}-r\left[\frac{\sigma^2}{2}+\gamma_{E}\right]$\\[1mm]\\[-2.5mm]
$\left(k\rightarrow 0\right)$ &  &  & \\[1mm]\hline\\[-2.5mm]
Rayleigh & $\ln\left\{c\,\mu_{r}\right\}-r\left[\frac{1}{\xi^{2}}+\frac{s^{2}}{2\,\sigma_s^{2}\,\xi^{2}\,\left(\xi^{2}+1\right)}+\ln\left\{\frac{\xi^2}{\xi^2+1}\right\}+\gamma_{E}\right]$ & $\ln\left\{c\,\mu_{r}\right\}-r\left[\frac{1}{\xi^{2}}+\ln\left\{\frac{\xi^2}{\xi^2+1}\right\}+\gamma_{E}\right]$ & $\ln\left\{c\,\mu_{r}\right\}-r\,\gamma_{E}$\\[1mm]\\[-2.5mm]
$\left(k\rightarrow 0; \sigma\rightarrow 0\right)$ &  &  & \\[1mm]\hline\\[-2.5mm]
M\'{a}laga ($\mathcal{M}$) & $\frac{r\,A\,\Gamma(\alpha)}{2^{r}}\sum_{m=1}^{\beta}b_{m}\,\Gamma(m)\,\left\{r\left[-1/\xi^2-\ln(B)-\frac{s^{2}\,\sigma_s^{-2}}{2\,\xi^{2}\,\left(\xi^{2}+1\right)}\right.\right.$ & $\frac{r\,A\,\Gamma(\alpha)}{2^{r}}\sum_{m=1}^{\beta}b_{m}\,\Gamma(m)\,\left\{r\left[-1/\xi^2-\ln(B)\right.\right.$ & $\frac{r\,A\,\Gamma(\alpha)}{2^{r}}\sum_{m=1}^{\beta}b_{m}\,\Gamma(m)\,\left\{r\left[-\ln(B)\right.\right.$\\[1mm]\\[-2.5mm]
 & $\left.\left.-\ln\left\{\frac{\xi^{2}}{\xi^{2}+1}\right\}+\psi(\alpha)+\psi(m)\right]+\ln(c\,\mu_{r})\right\}$ & $\left.\left.-\ln\left\{\frac{\xi^{2}}{\xi^{2}+1}\right\}+\psi(\alpha)+\psi(m)\right]+\ln(c\,\mu_{r})\right\}$ & $\left.\left.+\psi(\alpha)+\psi(m)\right]+\ln(c\,\mu_{r})\right\}$\\[1.2mm]\hline\\[-2.5mm]
Gamma-Gamma (GG) & $\ln\left\{c\,\mu_{r}\right\}-r\left[\frac{1}{\xi^{2}}+\frac{s^{2}}{2\,\sigma_s^{2}\,\xi^{2}\,\left(\xi^{2}+1\right)}\right.$ & $\ln\left\{c\,\mu_{r}\right\}-r\left[\frac{1}{\xi^{2}}+\ln\left\{\frac{\xi^{2}}{\xi^{2}+1}\right\}\right.$ & $\ln\left\{c\,\mu_{r}\right\}-r\left[\ln\left\{\alpha\,\beta\right\}-\psi\left(\alpha\right)-\psi\left(\beta\right)\right]$\\[1mm]\\[-2.5mm]
$\left(\rho\rightarrow 1,\,\Omega^{'}=1\right)$ & $\left.+\ln\left\{\frac{\xi^{2}}{\xi^{2}+1}\right\}+\ln\left\{\alpha\,\beta\right\}-\psi\left(\alpha\right)-\psi\left(\beta\right)\right]$ & $\left.+\ln\left\{\alpha\,\beta\right\}-\psi\left(\alpha\right)-\psi\left(\beta\right)\right]$ & \\[1.4mm]\hline\hline
\end{tabular}
\end{table}\normalsize
\end{landscape}

\end{document}